\documentclass[fleqn,usenatbib]{mnras}

\usepackage[T1]{fontenc}

\DeclareRobustCommand{\VAN}[3]{#2}
\let\VANthebibliography\thebibliography
\def\thebibliography{\DeclareRobustCommand{\VAN}[3]{##3}\VANthebibliography}

\usepackage{graphicx}	
\usepackage{amsmath}	
\usepackage{amssymb}	
\usepackage{ulem}
\usepackage{xcolor}

\newcommand\T{\rule{0pt}{2.6ex}}       
\newcommand\B{\rule[-1.2ex]{0pt}{0pt}} 

\title[Super-critical accretion on to seed BHs]{Super-critical
accretion of medium-weight seed black holes in gaseous proto-galactic nuclei}

\author[F. Sassano et al.]{Federica Sassano,$^{1,2}$\thanks{E-mail: \href{mailto:federica.sassano@uniroma1.it}{federica.sassano@uniroma1.it}} Pedro R. Capelo,$^{3}$ Lucio Mayer,$^{3}$  Raffaella Schneider$^{1,2,4}$ and Rosa Valiante$^{2,4}$
\\
$^{1}$Dipartimento di Fisica, Sapienza, Universit$\grave{a}$ di Roma, Piazzale Aldo Moro 5, 00185, Roma, Italy  \\
$^{2}$INFN, Sezione di Roma I, P.le Aldo Moro 2, 00185 Roma, Italy  \\
$^{3}$Center for Theoretical Astrophysics and Cosmology, Institute for Computational Science, University of Zurich,\\ Winterthurerstrasse 190,  CH-8057, Z{\"u}rich, Switzerland \\
$^{4}$INAF/Osservatorio Astronomico di Roma, Via di Frascati 33, 00078 Monte Porzio Catone, Italy}

\date{Accepted XXX. Received YYY; in original form ZZZ}

\pubyear{2022}

\begin{document}
\label{firstpage}
\pagerange{\pageref{firstpage}--\pageref{lastpage}}
\maketitle

\begin{abstract}
Accretion at sustained or episodic super-Eddington (SE) rates has been proposed as a pathway to grow efficiently light seeds produced by Pop-III stars. We investigate if SE accretion can be sustained onto a black hole (BH) with $M_{\rm BH} \sim 10^3$~M$_{\sun}$ in the centre of a gas-rich proto-galaxy at $z=15$. We perform high-resolution smoothed-particle hydrodynamical simulations, including two different sub-grid models for SE accretion, one based on the slim disc paradigm, and one inspired by recent radiation-magnetohydrodynamical simulations by Jiang and collaborators. Radiative feedback has the form of a thermal dump to surrounding gas particles, with the radiative efficiency being set according to the different SE accretion models. We find that, in all simulations, star formation, BH feedback, and interactions between clumps and the BH rapidly quench accretion after $\sim$1~Myr, irrespective of the sub-grid model used for accretion. Quenching is stronger in the model based on the simulations of Jiang and collaborators relative to the slim disc model because of its higher radiative efficiency. The SE growth phase is always very brief, lasting a few 0.1~Myr. In the most optimistic case, the BH reaches a mass of $\sim$10$^4$~M$_{\sun}$. We extrapolate the final BH masses from $z=15$ to $z\sim6$, assuming subsequent galaxy mergers will replenish the gas reservoir and trigger new cycles of SE accretion. We find that at most BH seeds would grow to $\sim$10$^6$~M$_{\sun}$, comparable to the mass of massive BHs in spiral galaxies such as the Milky Way, but falling short of the mass of the high-redshift quasars.
\end{abstract}

\begin{keywords}
quasars: supermassive black holes -- black hole physics -- accretion discs -- galaxies: evolution -- galaxies: high-redshift -- galaxies:nuclei.
\end{keywords}

\section{Introduction}

Observations of supermassive black holes (SMBHs) with $M_{\rm SMBH}\gtrsim 10^9$~M$_{\sun}$ at $z\geq 6$ \citep[e.g.][]{willott2010, Jiang2016, Banados2016, banados2018800,matsuoka2019discovery, yang2020,wang2021luminous} have opened a controversial debate on their possible origin \citep[][]{volonteri2012formation,inayoshi2020}. The greatest open question is related to the nature of their progenitor seeds, expected to form around the formation epoch of the first stars and black holes (BHs), at $z=20$--$30$ \citep[][]{alvarez2009accretion,bromm2013}. The first BH seeds are expected to originate from the collapse of the first metal-poor stars, or Pop-III stars. The remnant of such a massive star can be a \textit{light seed} of $\sim$100~M$_{\sun}$ \citep[][]{madau2001early, haiman2001highest,HegerWoosley02}.

Concerning the possibility of identifying light seeds as SMBH progenitors, several works show that unimpeded mass accretion rate on to a light seed is unrealistic for several reasons. First, because these BHs are expected to form in starvation, and the accretion process emits ionizing radiation that affects the gas supply on to the BH \citep[][]{park2012accretion}. Another reason is that the gas inflows through galactic filaments and mergers require relatively long time-scales ($\sim$100~Myr; \citealt{yoshida2008}), leading to an intermittent growth.

Some studies suggest the existence of more massive seeds, \textit{heavy seeds} with $\sim$10$^5$~M$_{\sun}$ originating from the direct collapse of supermassive stars formed in metal-poor regions irradiated by an intense photodissociating flux \citep[][]{omukai1998formation,omukai2008can,yoshida2008,lodato2006supermassive,lodato2007}, and {\it medium-weight seeds} with typical masses of $\sim$10$^3$~M$_{\sun}$, expected to form from the runaway collisions of stars in dense stellar cluster \citep[][]{volonteri2010formation,devecchi2012,katz2015,sakurai2017formation,stone2017formation,reinoso2018,tagawa2020making} or runaway mergers of BHs \citep[][]{davies2011supermassive,lupi2014constraining}. Their progenitor halos, pristine atomic cooling halos with virial temperatures $T_{\rm vir}\geq 10^4$~K, are expected to form at lower redshift compared to light seeds \citep[][]{dijkstra2014feedback,valiante2016first,habouzit2016number}. Even if heavy seeds could represent a good progenitor for a SMBH, they are extremely rare. On the contrary, medium-weight seeds, which are predicted to form in atomic cooling halos irradiated by an intense Lyman--Werner background but under more favorable conditions for the chemical enrichment \citep[][]{schneider2012,chon2020supermassive,sassano2021light}, could represent a viable BH seed candidate, particularly if super-Eddington (SE) accretion rates are considered.

Different authors have explored the possibility of SE accretion on to a massive BH seed \citep{volonteri2015case,lupi2016growing,pezzulli2016,pezzulli2017sustainable,regan2019super,toyouchi2021super}. Simulations from \citet{regan2019super} showed that mechanical feedback from bipolar jets, produced by accretion onto a $10^4$~M$_{\sun}$ BH, quickly evacuates high-density gas from the vicinity of the BH once the accretion rate exceeds the Eddington accretion rate. For this reason, accretion is rapidly quenched and the effective accretion rate drops to sub-Eddington rates on a relatively short time-scale ($\sim$0.1~Myr). Hence, they conclude that efficient growth can not be sustained if the host galaxy evolves in isolation, but may be restarted by gas inflows triggered by  galaxy mergers, as also suggested by \citet{pezzulli2016}. Recently, \citet{toyouchi2021super} investigated with 3D radiation-hydrodynamics simulations the efficient mass accretion onto a $10^4$~M$_{\sun}$ BH embedded in a massive, self-gravitating, dusty nuclear accretion discs. They found that the BH can be fed at rates exceeding the Eddington rate only when the dusty disc becomes optically thick to ionizing radiation, otherwise photoevaporation causes mass outflows from the disc strongly preventing BH feeding. These conclusions are sensitive to the anisotropy of the radiation field, since geometrical effects potentially change the accretion dynamics and the behaviour of the gas at small scales can significantly affect BH feedback \citep[][]{skadowski2016three,jiang2014global}.

\citet{sadowski2009,skadowski2016three} and \citet{jiang2014global,jiang2019super} have investigated the effects of mechanical and radiative feedback with two highly complex computational techniques. These two methods currently provide the most advanced numerical simulations and represent two different approaches to study SE accretion flows on to BHs of different masses with high accuracy and resolution. 

On the one hand, \citet{sadowski2009} studied the effect of feedback, improving slim disc solutions by performing fully relativistic, axisymmetric 2D hydrodynamical simulations. From these numerical solutions, an analytic formulation was provided by \citet[][]{madau2014}. These results have been successively confirmed also by their advanced 3D magnetohydrodynamical (MHD) simulations  \citep{skadowski2016three}. On the other hand, 3D radiation-MHD simulations by \citet[][]{jiang2014global,jiang2019super}, adopting different treatments of feedback, find a  more radiatively efficient disc compared to the slim-disc model described above.  

The purpose of this work is to study the regime of SE accretion on to a medium-weight seed in an  atomic cooling halo, adopting different treatments of feedback. 

In Section~\ref{section:cosmo}, we discuss and justify our choice for the initial conditions adopted to run our numerical simulations. In Section~\ref{section:modelGas}, we describe the model, the tree-smoothed particle hydrodynamics (SPH) $N$-body code \textsc{gasoline2} \citep[][]{wadsley2004gasoline,wadsley2017gasoline2}, and how we implement the two different treatments of feedback: the analytic recipes of \citet[][]{madau2014,sadowski2009} and  \citet{jiang2019super}. In Section~\ref{section:resultsGas}, we illustrate the results of our work, and compare these to recent independent studies  \citep{lupi2016growing,regan2019super,massonneau2022supereddington} in Section~\ref{section:SEdiscussion}. In Section~\ref{subsection:simplemodel}, we develop a simple model to extrapolate the evolution of a medium-weight BH seed and its host galaxy from $z=15$ to $z\sim 6$, assuming subsequent SE episodes rejuvenating the system. Finally, in Section~\ref{section:conclusions}, we present the main conclusions of our work.

\section{Cosmological constraints}\label{section:cosmo}

In this section, we present the properties of the environment that we adopt to describe the formation site of a newly formed medium-weight BH seed. These represent the initial conditions of the hydrodynamical simulation. In our recent work \citep{sassano2021light}, we  explored the formation rate of medium-weight BH seeds and its evolution in redshift based on the semi-analytical cosmological model \textsc{GQd} \citep[][]{valiante09,valiante2011origin,valiante2014high,valiante2016first,valiante2017,Valiante20,sassano2021light}. We investigated the physical conditions for light, medium-weight, and heavy BH seeds formation, exploring their birth rate within a self-consistent model. We found that the impact of the spatial fluctuation of the photo-dissociating radiation and inhomogeneities in the metal/dust enrichment are essential features for seed formation. We adopted different values for the critical photo-dissociating radiation threshold, $J_{\rm cr}$, to investigate the possible consequences on seed formation. We also implemented two different models: our reference model and a different model for massive seeds formation, the super competitive accretion (SCA) scenario  \citep[][]{chon2020supermassive}. As explained in \citet[][]{sassano2021light}, in the latter model a medium-weight BH seed can form  in slightly metal-enriched gas clouds ($10^{-3.5}$--$10^{-2.5}$~Z$_{\sun}$) illuminated by a strong ultraviolet field, with $J > J_{\rm cr}$, independently of the value of the dust-to-gas mass ratio. In fact, despite the fact that significant fragmentation occurs in a very compact region at the centre (100~AU) due to dust cooling, the latter does not affect the gas temperature on larger scales and the accretion rate remains very high, feeding the central primary star. In particular, we consider here the results obtained running the SCA300 model version, where $J_{\rm cr} = 3 \times 10^{23}$~erg~s$^{-1}$~Hz$^{-1}$~cm$^{-2}$~sr$^{-1}$.

\begin{figure}
    \centering
    \includegraphics [scale=0.16]{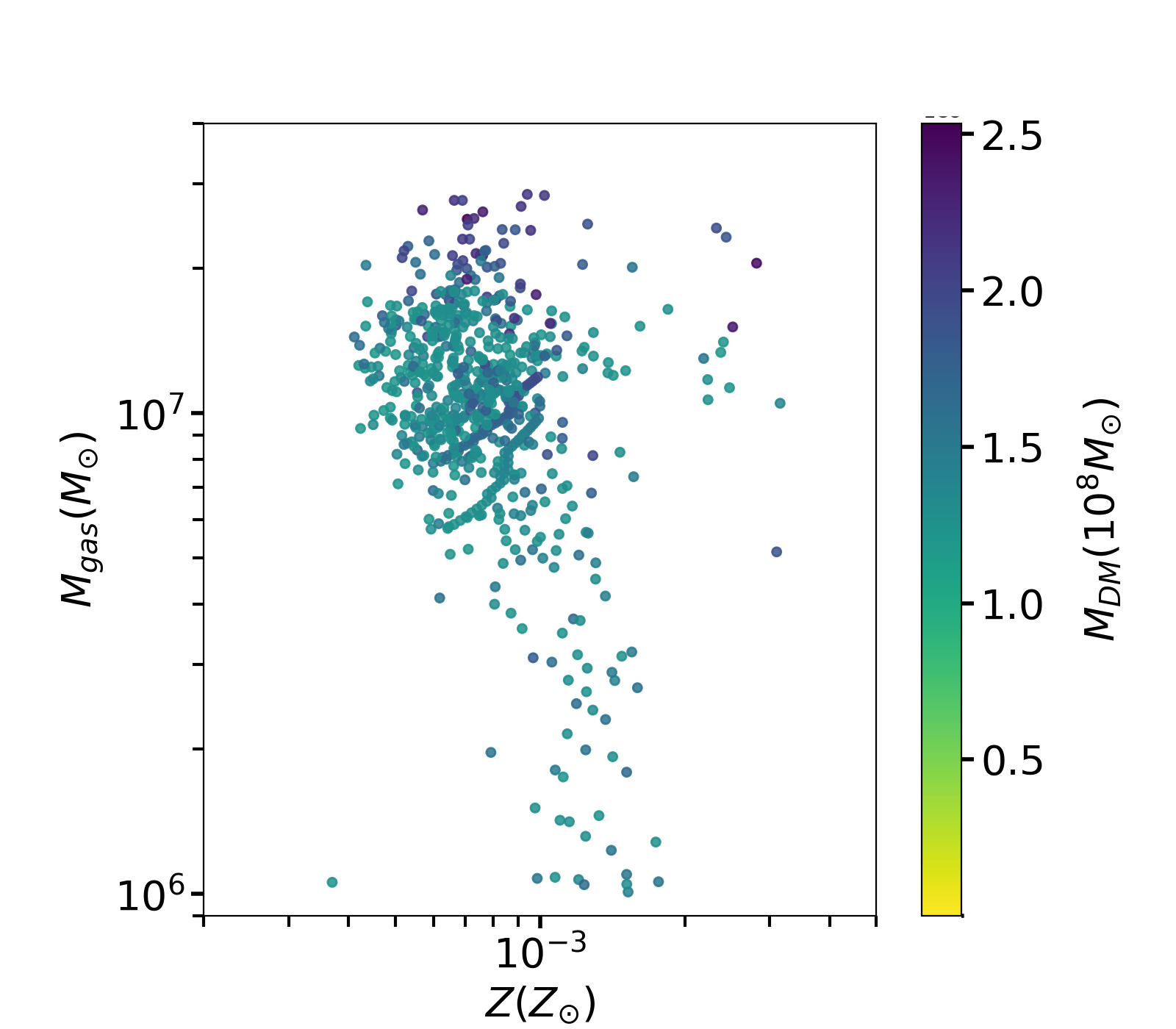}
    \caption{Gas mass as a function of gas metallicity and DM halo mass. Each point represents the host of a medium-weight seed forming in the redshift range $14.5<z<15.5$, predicted by \textsc{GQd} (see text).}
    \label{fig:sem}
\end{figure}

Fig.~\ref{fig:sem} shows the gas mass as a function of metallicity and dark matter (DM) halo mass, in galaxies hosting a newly formed medium-weight BH in the redshift range $14.5 < z < 15.5$. Each of the $\sim$670 points represents an individual system, as predicted by one SCA300 simulation of \textsc{GQd} \citep{sassano2021light}. In this redshift range, the metallicity of the hosts varies from $\sim$4$\times10^{-4}$ to $\sim$2$\times 10^{-3}$~Z$_{\sun}$, and a large number of seeds form in halos with DM mass of $\sim$10$^8$M$_{\sun}$ and gas mass of $\sim$10$^7$~M$_{\sun}$.

\subsection{Initial conditions}

Based on our previous findings, we adopt the following initial conditions for the hydrodynamical simulation: we assume that a medium-weight BH seed of $10^3$~M$_{\sun}$ is formed in the inner region of a DM halo of mass $M_{\rm DM}=2.5\times 10^7$~M$_{\sun}$ at $z = 15$, which we sample with $N_{\rm DM} = 10^6$ particles. This choice enables us to reach a DM particle mass resolution of 25~M$_{\sun}$, allowing us to properly describe the dynamical effects acting on the central $10^3$~M$_{\sun}$ BH with a limited computational cost. This is a necessary condition if we want to perform a set of simulations to explore the effects of different BH accretion and feedback models. The adopted total DM mass is however smaller than the mass of halos that host the formation of medium-weight BHs (see Fig.~\ref{fig:sem}). Our simulation is therefore targeting the inner region encompassing $\sim$25 per cent of the total DM mass of the host halo. We further assume that the gas mass of the central region is $5 \times 10^6$~M$_{\sun}$, $\sim$50~per cent of the total gas mass of a typical host halo, and the gas metallicity is $10^{-3}$~Z$_{\sun}$. In our model, we assume that no star formation (SF) occurs concurrently in the halo when a medium-weight seed forms: initially, no stars are present in the disc. 

\begin{figure}
    \centering
    \includegraphics [scale=0.36]{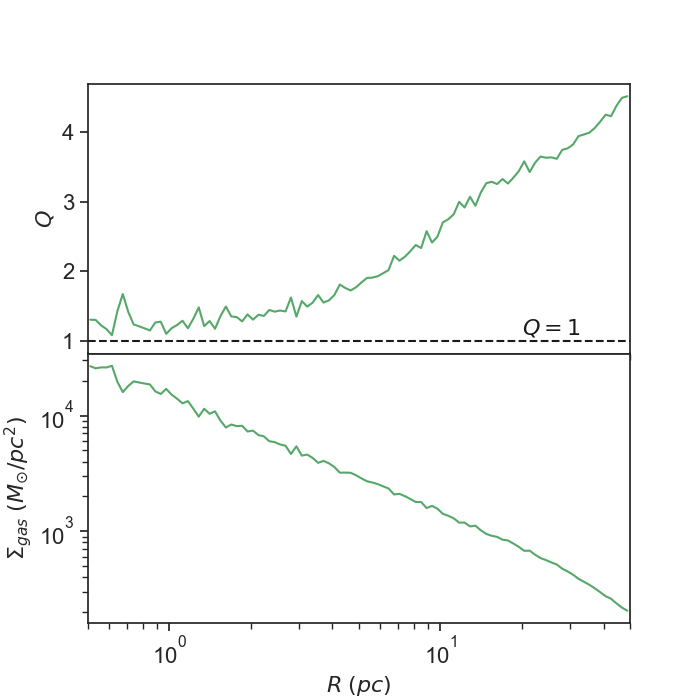}
    \caption{Initial Toomre parameter $Q$ (top panel) and gas surface density $\Sigma_{\rm gas}$ (bottom panel) as a function of the radial distance from the centre.} 
    \label{fig:Incond}
\end{figure}

An appropriate set of initial conditions are selected in order to describe the evolution of this system under the effects of gas cooling, SF, and feedback from the accreting BH. To choose the initial radii of the inner gas and DM regions of our proto-galaxy, $R_{\rm Gas}$ and $R_{\rm DM}$, we assume that for $z>0$ the two radii decrease approximately as $(1+z)^{-1/3}R_{z=0}$, where $R_{z=0}$ represents the typical local value. In our choice we are guided by the work of \citet{SouzaLima2017,SouzaLima2020}, who studied BH accretion and feedback in a circumnuclear disc (CND) embedding a BH pair of masses $5 \times 10^5$~M$_{\sun}$ and $10^7$~M$_{\sun}$ at $z=0$. We model the DM density profile as a \citet{Plummer_1911} sphere with a Plummer radius $r_{\rm P}=18$~pc (and we also assume a maximal radius $R_{\rm DM} \sim 184$~pc for numerical reasons). The gas component is distributed as a \citet{Mestel_1963} disc with a disc maximum radius $R_{\rm gas} = 55$~pc and scale height $H/R = 0.05$. The initial radial profile of the gas surface density, $\Sigma_{\rm gas}(R)$, is shown in the bottom panel of Fig.~\ref{fig:Incond}. The disc, of mass $5 \times 10^6$~M$_{\sun}$, is sampled by $2\times 10^5$ particles, so that the mass of a baryonic particle is 25~M$_{\sun}$, the same as the DM particles mass. The gravitational softening length is $r_{\rm soft} = 0.18$~pc for all particles and the initial gas temperature and metallicity are assumed to be $T_0 = 2500$~K, to have an initially Toomre-stable disc (\citealt{Toomre_1964}; see the top panel of Fig.~\ref{fig:Incond}), and $Z = 10^{-3}$~Z$_{\sun}$ (see Fig.~\ref{fig:sem}).

\begin{table*}
\caption{Comparison between the initial conditions adopted in our work and in  that of \citet{SouzaLima2017}.} 
\begin{tabular}{|c|ccccccc|}
\hline
\hline
Work & $M_{\rm DM}$ & $R_{\rm DM}$ & $N_{\rm DM}$ &  $M_{\rm gas}$&  $R_{\rm gas} $ & $N_{\rm gas} $  & $r_{\rm soft}$  \T \B \\
\hline
\hline
\citet[][]{SouzaLima2017} & $5 \times 10^8$~M$_{\sun}$ & 500~pc & $10^6$ & $10^8$~M$_{\sun}$ & 150~pc & $2 \times 10^5$ & 0.5~pc  \T \B \\
\hline
This work                 & $2.5 \times 10^7$~M$_{\sun}$ & 184~pc & $10^6$ & $5 \times 10^6$~M$_{\sun}$ & 55~pc & $2 \times 10^5$ & 0.18~pc  \T \B \\
\hline
\hline
\end{tabular}
\label{tab:tableIC}
\end{table*}

\section{Numerical simulations}\label{section:modelGas}

We run a set of numerical simulations adopting the tree-SPH $N$-body code \textsc{gasoline2} \citep[][]{wadsley2004gasoline,wadsley2017gasoline2}. \textsc{gasoline2} was built on the \textsc{pkdgrav} $N$-body code \citep[][]{stadel2001cosmological}, which assumes pure gravity, by adding hydrodynamics with the SPH method, wherein gas particles are the fundamental resolution elements (rather than grid cells), and several sub-grid recipes that model sub-resolution physical phenomena, such as SF and BH feedback. Here we briefly summarize some important features of this model.

\subsection{Gas cooling}

The model assumes photoionization non-equilibrium cooling due to the following primordial species: H, H$^+$, He, He$^+$, and He$^{++}$. We do not consider H$_2$. This can be partly justified by the peculiar conditions required to form a medium-weight BH seed, in which molecular cooling is inefficient due to H$_2$ photo-dissociation by the Lyman--Werner radiation field.

Metal cooling affects the gas temperature, ionization states, and gas dynamics. The photoionization equilibrium cooling from metals is that provided by \citet{shen2010enrichment,shen2013circumgalactic}, who computed the cooling rate for the solar case and presented a simple rescaling for non-solar abundances, according to which the cooling rate $\Lambda(Z)=(Z/{\rm Z}_{\sun})\Lambda({\rm Z}_{\sun})$. We set the temperature cooling floor to $T_{\rm min} \sim 44$~K, which corresponds to the temperature of the cosmic microwave background at the simulated redshift ($z = 15$); this ensures that the \citet{Jeans_1902} mass is always resolved by at least 64 gas particles for all densities above the SF density threshold (see below).

\subsection{Star formation}

SF can occur in dense ($> 1000$ cm$^{-3}$) and cold ($< 10^3$~K) gas particles \citep[][]{stinson2006star}. A subset of these gas particles are stochastically selected \citep[][]{katz1992dissipational} to form stars according to

\begin{equation}
    p= \frac{m_{\rm gas}}{m_{*}}(1-e^{-\epsilon_{\rm SF}\Delta t/ \tau_{\rm SF}}),
\end{equation}

\noindent where $m_{\rm gas}$ and $m_{*}$ are the gas and star particle masses, respectively (in our case, $m_{\rm gas} = m_{*}$), $\Delta t = 0.01$~Myr is how often SF is computed, $\epsilon_{\rm SF} = 0.05$ is the SF efficiency, and $\tau_{\rm SF}$ is the local dynamical time, yielding a SF rate (SFR) \citep{schmidt1959rate}

\begin{equation}
    \frac{{\rm d}\rho _*}{{\rm d}t}=\epsilon_{\rm SF} \frac{\rho_{\rm gas}}{\tau_{\rm SF}},
\end{equation}

\noindent where $\rho_{\rm gas}$ is the gas density.

The code implements also stellar feedback, in the form of supernovae and stellar winds \citep[see][]{stinson2006star}. However, given the short duration of the runs, these events do not occur in our simulations.

\subsection{Black hole accretion}

\textsc{gasoline2} uses sub-grid models for BH accretion and feedback, since BH accretion scales require a huge dynamical range, ranging from $\mu$pc scales (the Schwarzschild radius of a SMBH) up to kpc scales (disc size). The most commonly used sub-grid accretion models are based on the Bondi--Hoyle--Lyttleton \citep[hereafter Bondi;][]{Hoyle_Lyttleton_1939,bondi1944mechanism,bondi1952spherically} accretion recipe. The Bondi accretion rate in the unresolved region is defined as

\begin{equation}
 \dot m=\alpha 4 \pi G^2 M_{\rm BH}^2 \frac{\rho_{\rm gas}}{(V^2+c_{\rm s}^2)^{3/2}},
 \label{eq:bondi}
\end{equation}

\noindent where $G$ is the gravitational constant, $M_{\rm BH}$ is the BH mass, $V$ is the local velocity of the gas with respect to the BH, $c_{\rm s}$ is the local speed of sound, and $\alpha$ is the boost factor (we adopt a boost factor $\alpha=1$, as in \citealt{SouzaLima2017}), used to correct for the lack of resolution, effectively allowing the gas density around the BH to be larger than predicted by the simulation, and that sometimes has been assumed to depend on the equation of state of the gas \citep{booth2009cosmological}.
 
In \textsc{gasoline2}, a BH is treated as a sink particle. The accretion rate is computed as the sum of the Bondi accretion rate of each of the 64 neighbour gas particles. Mass is then removed from the gas particle proportionally to their contribution, and added to the BH. Only in the Eddington-capped case, hereafter the standard case (STD), the accretion rate is capped at the standard Eddington mass accretion rate,

\begin{equation}
    \dot m_{\rm E,S}=\frac{L_{\rm E}}{\epsilon_{\rm r} c^2},
    \label{eq:medd}
\end{equation}

\noindent where $\epsilon_{\rm r} = 0.1$ is the radiative efficiency and $L_{\rm E}$ is the Eddington luminosity, defined as

\begin{equation}
    L_{\rm E} = \frac{4\pi G M_{\rm BH}c}{\mu_{\rm e} m_{\rm p} \sigma_{\rm T}},
    \label{eq:LEdd}
\end{equation}

\noindent with $\mu_{\rm e}$ being the mean molecular weight per electron, $m_{\rm p}$ the proton mass, $\sigma_{\rm T}$ the Thomson scattering cross-section, and $c$ the speed of light.

The net mass growth of the BH is given by

\begin{equation}
    \dot M_{\rm BH} = (1-\epsilon_{\rm r})\dot m,
    \label{eq:dmacc}
\end{equation}

\noindent where $\dot m$ is the Bondi mass accretion rate (see Eq.~\ref{eq:bondi}). In the present study, when not considering the standard case, we assume the radiative efficiency to be a function of the ratio between the Bondi accretion rate and $L_{\rm E}/c^2$, and to depend on the model adopted, as we explain in the following sections. However, it is important to stress that the physics of accretion discs supporting SE flows is still debated. Most models rely on a dominance of advection over radiation in transporting energy through the disc, in order to support a radiative efficiency significantly lower than in Eddington-limited flows, thus allowing to accrete at much higher rates without being impeded by radiation pressure. One of the most popular models is the slim disc model \citep{abramowicz1988slim,sadowski2009}. Fully 3D radiation-MHD simulations \citep[][]{skadowski2016three}, however, show that the properties of accretion discs can depart significantly from the assumptions in the slim disc model, because of MHD-induced turbulent transport and the complex geometry of the radiation field \citep[see the review by][]{mayer2019super}.

\subsection{Black hole feedback}

BH feedback in \textsc{gasoline2} is described in the form of the so-called thermal or radiative feedback: the radiated luminosity is a fraction $\epsilon_{\rm r}$ of the rest energy per unit time of the gas accretion rate and a small fraction of this, $\epsilon_{\rm c}$, is assumed to thermally couple to the gas. Hence, the rate of thermal energy released to the gas by BH accretion is computed as

\begin{equation}
    \dot E_{\rm c}=  \dot m \epsilon_{\rm r} \epsilon_{\rm c} c^2
    \label{eq:Efb}
\end{equation}

\noindent and it is distributed over the SPH kernel as thermal energy to the 64 particles nearest the BH \citep{bellovary2013}, and $\epsilon_{\rm FB}=\epsilon_{\rm r} \epsilon_{\rm c}$ is the BH feedback efficiency. We discuss below the values adopted for the two parameters $\epsilon_{\rm r}$ and $\epsilon_{\rm c}$. Here we remark that -- although the increased thermal energy of the gas surrounding the BH may eventually drive a gas outflow -- we do not consider the possibility that accreting BHs may exert mechanical feedback by powering the formation of bipolar jets. Although these jets may be highly collimated, recent simulations in the context of SE accretion show that they can significantly limit BH accretion by efficiently evacuating the region surrounding the BH \citep{regan2019super}. We will return to this point in Section~\ref{section:SEdiscussion}.

\subsection{Slim disc model}

We have implemented the effect of BH feedback following two different models for the radiative efficiency. In the first model, $\epsilon_{\rm r}$ is described according to \citet[][]{madau2014}, who provide an analytic fit of numerical solutions of the relativistic slim accretion disc equations of \citet[][]{sadowski2009}. Despite SE accretion rates, slim discs remain only moderately luminous, as a large fraction of the viscosity-generated heat is advected inward and released closer to the BH or not released at all. As a result of the increasing rate of advection, the efficiency of transforming gravitational energy into radiative flux decreases with increasing accretion rate. The fit of the bolometric luminosity $L$ provided by \citet{madau2014} is

\begin{equation}
L/L_{\rm E}= A(a)\biggr[ \frac{0.985}{\dot m_{\rm E,M}/\dot m + B(a)} + \frac{0.015}{\dot m_{\rm E,M}/\dot m + C(a)} \biggl],
\label{eq:lmadau}
\end{equation} 

\noindent where $\dot{m}_{\rm E,M} \equiv 16 L_{\rm E}/c^2$ and the functions A, B, and C,

\begin{equation}
A(a) = (0.9663-0.9292a)^{-0.5639},
\end{equation}

\begin{equation}
B(a) = (4.627-4.445a)^{-0.5524},
\end{equation}

\begin{equation}
C(a) = (827.3-718.1a)^{-0.7060},
\end{equation}

\noindent depend on the spin parameter of the BH, defined as $a = J/(cM_{\rm BH}^2)$, where $J$ is the BH angular momentum. The  radiative efficiency is then assumed to be $\epsilon_{\rm r,M} = 1/16 (L/L_{\rm E}) \,(\dot m_{\rm E,M}/ \dot m)$.

One important aspect is the dependence on the spin: for increasing values of the spin, the radiative efficiency increases as well. However, even when the spin is very high, e.g. $a=0.99$, in the super-Eddington regime the radiative efficiency can be one order of magnitude lower that the commonly adopted value of $\epsilon_{\rm r}=0.1$. In addition to the reduced effect of BH feedback, the BH mass growth also increases, as $\dot M_{\rm BH}=(1-\epsilon_{\rm r})\dot m$. The results show that the slim disc remains moderately luminous despite SE accretion rates. For higher values of $\dot m$, the disc becomes radiatively inefficient and $\epsilon_{\rm r}$ decreases.

\begin{figure}
    \hspace{-1 cm}
    \centering
    \includegraphics [scale=0.34]{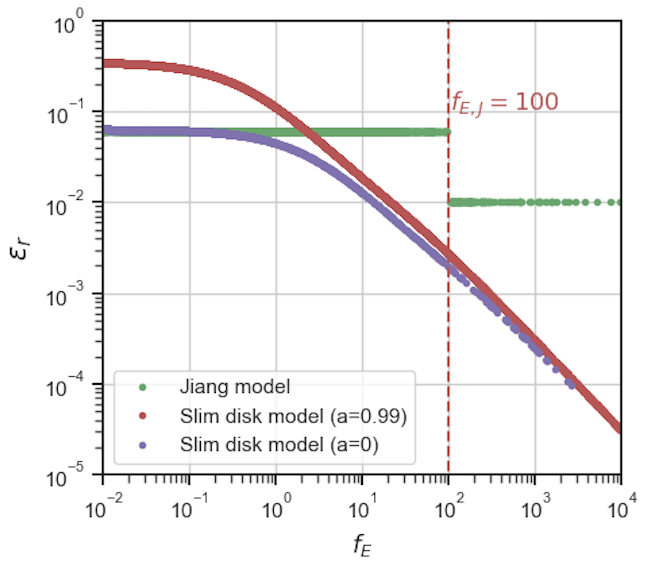}
    \caption{Radiative efficiency sampled as a function of the Eddington ratio, $f_{\rm E} = \dot m/\dot m_{\rm E}$, in the Jiang model (green dots), and slim disc models with two values of the spin, $a=0$ (purple dots), and $a=0.99$ (red dots). In the Jiang model, a non-spinning BH is considered. In order to compare different models, we always assume the corresponding definition of $\dot{m}_{\rm E}$, i.e. $\dot{m}_{\rm E}=\dot{m}_{\rm E,M}$ and $\dot{m}_{\rm E}=\dot{m}_{\rm E,J}$ for the slim disc and Jiang model, respectively.}
    \label{fig:jiangmadau}
\end{figure} 

In Fig.~\ref{fig:jiangmadau}, we report the radiative efficiency, $\epsilon_{\rm r}$, as a function of the Eddington ratio, $f_{\rm E}$ (i.e. the ratio between the BH accretion rate and the Eddington accretion rate), for the \citet{jiang2014global} and \citet{madau2014,sadowski2009} models (hereafter Jiang and slim disc models). For the slim disc model, there is a dependence on the BH spin when $(\dot m/ \dot m_{\rm E,M}) \lesssim 1$ and we show the radiative efficiency assuming a non-spinning BH (purple dots) and a very high value of $a=0.99$ (red dots). Following \citet{lupi2016growing}, we select $a = 0.99$ as our reference slim disc case, but we will also discuss the implications of adopting the radiative efficiency predicted for a non-spinning BH.

\subsection{Jiang model}

The second model that we explore is based on the results of  simulations performed by \citet[][]{jiang2014global,jiang2019super}. For their study, they adopt pseudo-Newtonian 3D MHD simulations, to simulate the effects of general relativity around a non-spinning BH, with enhanced radiative transport schemes, recently applied in the study of a $\sim$10$^8$~M$_{\sun}$ SMBH. In this case, different feedback treatments provide  different results with respect to the slim disc model discussed above. The main differences are probably due to the higher anisotropy in the propagation of photons and the turbulent transport. According to \citet[][]{jiang2014global,jiang2019super}, turbulence works in addition to advection, helping the photons propagate outwards. As a result, the radiative efficiencies found by \citet[][]{jiang2019super} in the SE regime are larger than those expected by the \citet{madau2014} slim disc model. In particular they run a set of MHD simulations for a non-spinning SMBH of $5 \times 10^8$~M$_{\sun}$, finding that $\epsilon_{\rm r,J}=0.06$ for $\dot m/\dot m_{\rm E,J}<100$, and $\epsilon_{\rm r,J}=0.01$ for $\dot m/\dot m_{\rm E,J}>100$ (see Fig.~\ref{fig:jiangmadau}), where $\dot{m}_{\rm E,J} \equiv 10 L_{\rm E}/c^2$ and, for consistency, $\epsilon_{\rm r,J} = 1/10 (L/L_{\rm E}) \,(\dot m_{\rm E,J}/ \dot m)$. They find that magnetic buoyancy allows photons to escape from the vertical direction of the disc before being advected into the BH. As a consequence, the radiative efficiency is less sensitive to large accretion rates.\footnote{When comparing different models, we will assume for each model its own definition of $\dot{m}_{\rm E}$, i.e. $\dot{m}_{\rm E,S}$, $\dot{m}_{\rm E,J}$, or $\dot{m}_{\rm E,M}$.}

It is important to stress that, in this model, the strength of the outflows is more pronounced with respect to \citet[][]{madau2014}, as a consequence of the more efficient radiative transport. However, due to the different nature of the radiative transport, the main reasons of these discrepancies are still unclear \citep[see section 2.1 of][]{mayer2019super}.

Throughout this work, we assume a model-dependent radiative efficiency $\epsilon_{\rm r}=[\epsilon_{\rm r,M},\epsilon_{\rm r,J}]$, and we will perform two sets of simulations, adopting both the slim disc and Jiang prescriptions for the dependence of the radiative efficiency on the BH accretion rate, as described above.

\subsection{Coupling efficiency}

\begin{table*}
\centering
\caption{Information on the set of simulations that we have performed: name of the run, model, radiative efficiency, coupling efficiency, and BH mass at $t = 1$~Myr (except for run M1e-5, for which we provide the value at 0.9~Myr), in units of $10^3$~M$_{\sun}$. The initial BH mass is $10^3$~M$_{\sun}$ in all simulations.}
\hspace{0.43 cm}
\begin{tabular}{|cccc|c|}
\hline
  Name & Model & $\epsilon_{\rm r}$ & $\epsilon_{\rm c}$& $M_{\rm BH,\sim1\,Myr}(10^{3}\rm M_{\sun})$ \T \B \\
  \hline
  M1e-3 & SE slim disc $(a=0.99)$ & $\epsilon_{\rm M}$ & $10^{-3}$ & $1.68$ \T \B \\
  M1e-5 & SE slim disc $(a=0.99)$ & $\epsilon_{\rm M}$ & $10^{-5}$  & $23.6$ \T \B \\
  M1e-6 & SE slim disc $(a=0.99)$ & $\epsilon_{\rm M}$ & $10^{-6}$ &  $20.3$ \T \B \\
  \hline
  J1e-3 & SE Jiang & $\epsilon_{\rm J}$ & $10^{-3}$ &  $1.22$ \T \B \\
  J1e-5 & SE Jiang & $\epsilon_{\rm J}$ & $10^{-5}$ &  $2.91$ \T \B \\
  J1e-6 & SE Jiang & $\epsilon_{\rm J}$ & $10^{-6}$ &  $18.6$ \T \B \\
  \hline
  STD1e-3 & Eddington-capped & $0.1$ & $10^{-3}$&  $1.009$ \T \B \\
  STD1e-5 & Eddington-capped & $0.1$ & $10^{-5}$ & $1.004$ \T \B \\
 \hline
  Ma01e-3 & SE slim disc $(a=0)$ & $\epsilon_{\rm M}$ & $10^{-3}$ & $1.30$ \T \B \\
  Ma01e-5 & SE slim disc $(a=0)$ & $\epsilon_{\rm M}$ & $10^{-5}$ & $18.2$ \T \B \\
  Ma01e-6 & SE slim disc $(a=0)$ & $\epsilon_{\rm M}$ & $10^{-6}$ &  $19.9$ \T \B \\
   \hline
  NOFB & SE Jiang & $\epsilon_{\rm J}$ & 0.0 & $20.3$ \T \B \\
  \hline
\end{tabular}
\label{tab:table1Gas}
\end{table*}
 
The coupling efficiency, $\epsilon_{\rm c}$, represents the fraction of energy released by the BH that thermally couples to the gas \citep[e.g.][]{van2014nuclear,Capelo2015}. In cosmological simulations, this parameter is usually calibrated in order to reproduce the locally observed scaling relation between the mass of the nuclear BH and the mass of the stellar bulge \citep[][]{dimatteo2005energy, bellovary2013}.

The studies of \citet[][]{skadowski2016three} and \citet{jiang2019super} show that the geometry of the system can significantly alter the ratio between radiative and mechanical energy. In particular, more collimated jets can enhance mechanical feedback, lowering the strength of radiative feedback. This complex interplay among different feedback mechanisms could significantly decrease the coupling efficiency $\epsilon_{\rm c}$. For this reason, we vary this parameter in our simulations, to explore its impact on different models. In the future, we aim to refine this description by considering the dependence of $\epsilon_{\rm c}$ on $\epsilon_{\rm r}$, and eventually on the mass accretion rate. Here we explore different values for the coupling efficiency in the range $10^{-3}$--$10^{-6}$, and we also run the case without feedback (i.e. $\epsilon_{\rm c} = 0$). The goal is to test the effects of the two SE models on the early mass growth of BH seeds considering different values for the coupling efficiency.

\subsection{Simulation runs}\label{subsection:Simulation runs}

The list of all runs we performed is presented in Table~\ref{tab:table1Gas}. For both models (Jiang and slim disc), we explored three different values of $\epsilon_{\rm c}$ -- $10^{-3}$, $10^{-5}$, $10^{-6}$ -- and we named the corresponding simulations as J1e-3, J1e-5, J1e-6 and M1e-3, M1e-5, M1e-6, respectively. For completeness, we also considered a model without BH feedback, assuming $\epsilon_{\rm c} = 0$. These choices of $\epsilon_{\rm c}$ are motivated by the attempt to simulate different modes of mechanical feedback, wherein lower values of $\epsilon_{\rm c}$ correspond to even more collimated bipolar jets. The default value $\epsilon_{\rm c}=10^{-3}$ is the value commonly adopted in this kind of simulations \citep[e.g.][]{Capelo2015,SouzaLima2017}. Then we progressively reduce this value by 1 dex, until Jiang and slim disc models show similar behaviour (i.e. when $\epsilon_{\rm c} = 10^{-6}$), when the coupling is reduced to the point at which feedback by the BH does not affect significantly the mass growth. We report in addition the case without feedback, $\epsilon_{\rm FB}=\epsilon_{\rm c}=0$, although we will not discuss its results in detail and we will present only the most representative runs among the whole set that we have performed: the slim disc and Jiang models with $\epsilon_{\rm c}=10^{-3}$, $10^{-5}$, and $10^{-6}$. In addition, we performed two runs, STD1e-3 and STD1e-5, where we impose a (standard) Eddington-limited growth (with $\epsilon_{\rm r} = 0.1$) and adopt two different values for the coupling efficiency, $\epsilon_{\rm c}=10^{-3}$ and $10^{-5}$. These additional runs will not be discussed in detail, but they will be mentioned in comparison with SE models. Finally, for completeness, in the Appendix we present the results of two additional simulations of the slim disc model (Ma01e-3, and Ma01e-6), assuming a non-spinning BH ($a = 0$) and two values of the coupling efficiency, $\epsilon_{\rm c}=10^{-3}$, and $10^{-6}$.

\section{Results}\label{section:resultsGas}

In this section, we discuss the main results and the differences between the Jiang and slim disc models for different coupling efficiencies. In particular, we focus on BH accretion and how the final BH mass is affected by adopting different feedback models. 

\subsection{Black hole evolution}

Here we discuss the time evolution of the BH mass for all models. We start at $t = 0$, adopting an initial BH mass of $10^3$~M$_{\sun}$. The values of the BH mass at $t \sim 1$~Myr are reported in the last column of Table~\ref{tab:table1Gas}. 

\begin{figure*}
    \centering
    \includegraphics [scale=0.5]{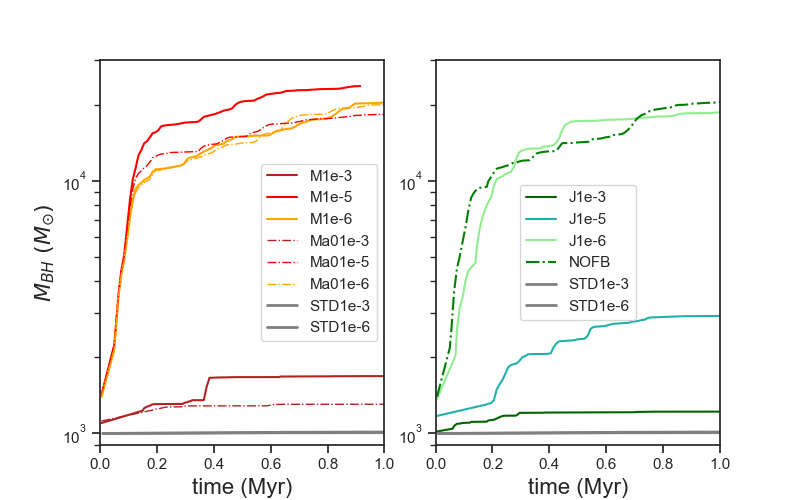}
    \caption{BH evolution as a function of time. Left- and right-hand panels show, respectively, results from the slim disc and Jiang models. Grey lines in both panels represent Eddington-limited models.}
    \label{fig:bhevo}
\end{figure*}

The duration of the simulation is not the same in all runs. Since we find that the mass of the BH saturates at $\sim$1~Myr, we ensured that all simulations (except for one) reach 1~Myr. Due to computational costs, the shortest run, M1e-5, lasts only $t_{\rm max} = 0.9$~Myr, but by this time the BH mass growth has already saturated in this model.

The evolution of the BH mass as a function of time is plotted in Fig.~\ref{fig:bhevo}. We find that BH evolution depends on the feedback model adopted. In general, lower coupling efficiencies provide higher BH masses, as expected. In the slim disc models, the final BH mass varies from $1.68 \times 10^3$~M$_{\sun}$ for $\epsilon_{\rm c} = 10^{-3}$ to $20.3 \times 10^3$~M$_{\sun}$ for $\epsilon_{\rm c} = 10^{-6}$ (and a similar value is found for $\epsilon_{\rm c} = 10^{-5}$). In the Jiang models, BH growth is less efficient with respect to the slim disc models. The mass growth in J1e-3 and J1e-5 is similar, with the final BH mass being, respectively, $1.22 \times 10^3$~M$_{\sun}$ and $2.91 \times 10^3$~M$_{\sun}$, whereas it suddenly increases when $\epsilon_{\rm c} = 10^{-6}$. The final BH mass in this case is $18.6 \times 10^3$~M$_{\sun}$, comparable to the results obtained in M1e-5 and M1e-6, as well as in the no feedback model.\footnote{The final BH mass in the run M1e-5 is slightly higher than that in M1e-6, even though it has a larger coupling efficiency. This is due to the very low values of the feedback efficiency (of the order of $10^{-8}$) in both runs, which make the chaotic gas structure more relevant.}

When a non-spinning BH is considered for the slim disc models (runs Ma01e-3, Ma01e-5, and Ma01e-6), the final BH mass ranges from $1.3 \times 10^3$~M$_{\sun}$ to $19.92 \times 10^3~$M$_{\sun}$. We will discuss the differences with the maximally spinning BH models in the Appendix, focusing in particular on models Ma01e-3 and Ma01e-6.

For comparison, we also show the mass evolution in the standard models, STD1e-3 and STD1e-5, wherein BH accretion is Eddington-limited (and $\epsilon_{\rm r} = 0.1$). We find that the final BH mass is higher when we adopt a lower coupling efficiency, $\epsilon_{\rm c}=10^{-5}$. However, BH growth is quenched in both models and only gains 1--4 per cent of the initial mass.  

In conclusion, super-Eddington growth leads to larger BH masses compared to the standard, Eddington-limited model, and the mass growth in the slim disc models is more sensitive to feedback than in the Jiang models (note the difference between the M1e-5 and J1e-5 models). A detailed comparison between the two models is presented below.

\begin{figure*}
    \centering
    \includegraphics [scale=0.6]{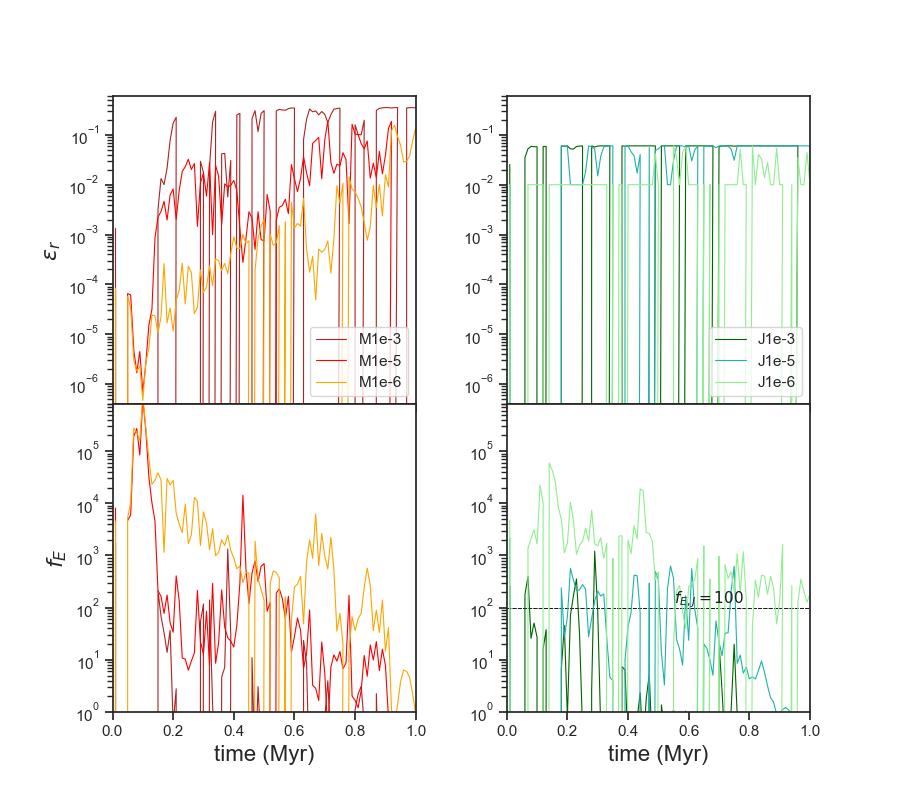}
    \caption{Time evolution of the radiative efficiency (top panels) and Eddington mass accretion ratio (bottom panels), for the Jiang (right-hand panels) and slim disc models (left-hand panels), adopting a coupling efficiency of $10^{-3}$ (J1e-3, M1e-3), $10^{-5}$ (J1e-5, M1e-5), and $10^{-6}$ (J1e-6, M1e-6). The horizontal dashed line in the bottom-right panel indicates $f_{\rm E,J}=100$.
    }
    \label{fig:1e3}
\end{figure*}

\begin{figure*}
    \centering
    \includegraphics [scale=0.24]{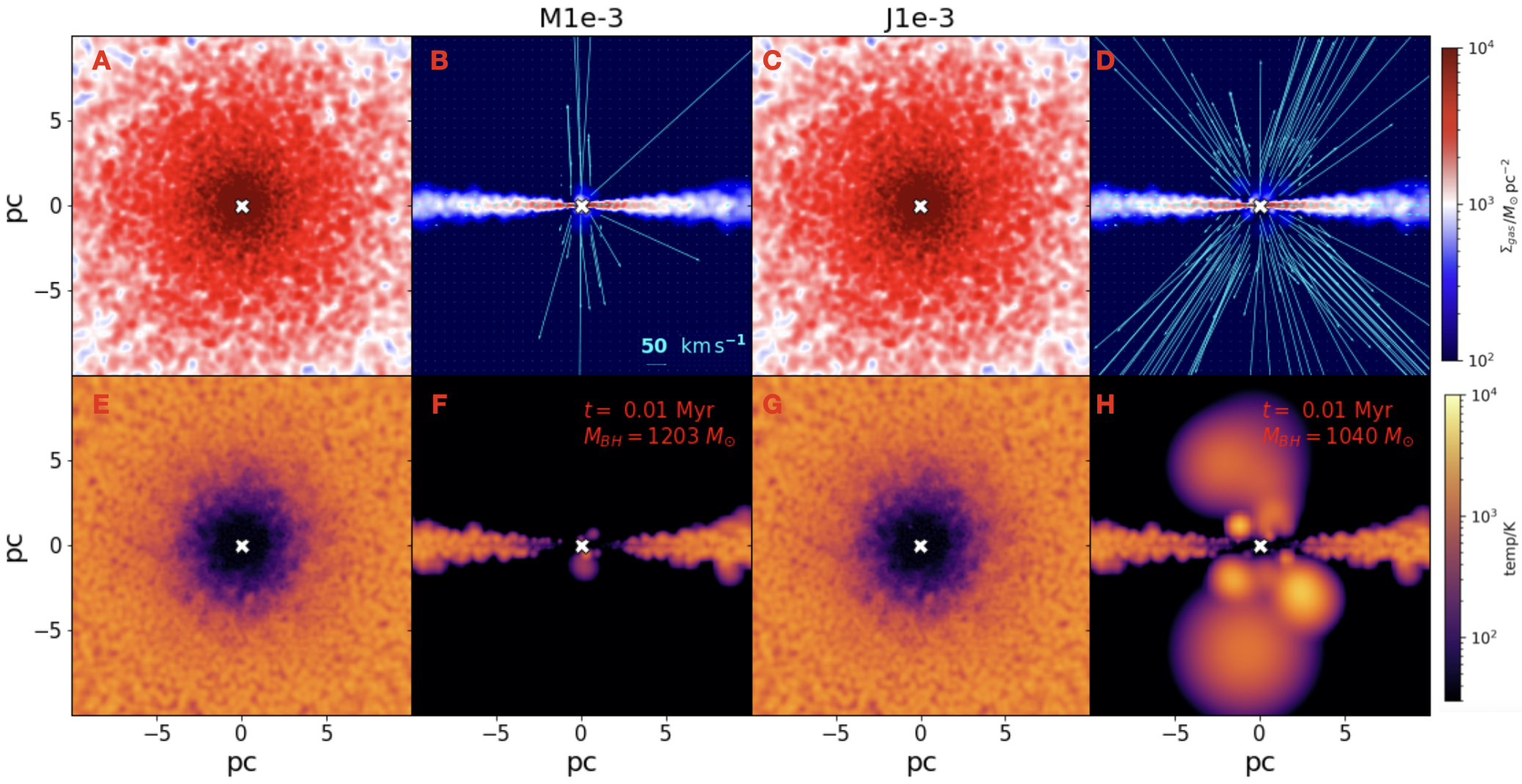}
    \caption{Gas surface density (top panels) and mass-weighted temperature (bottom panels)  maps of the disc at $t=0.01$~Myr, for the two runs M1e-3 (panels A, B, E, and F) and J1e-3 (panels C, D, G, and H). Odd and even columns show a face-on and edge-on view of the disc, respectively. Cyan arrows indicate the direction and magnitude of the gas velocity flows, normalized at the corresponding value. The white crosses mark the positions of the BH.}
    \label{fig:snap1e3}
\end{figure*}

\begin{figure*}
    \centering
    \includegraphics [scale=0.26]{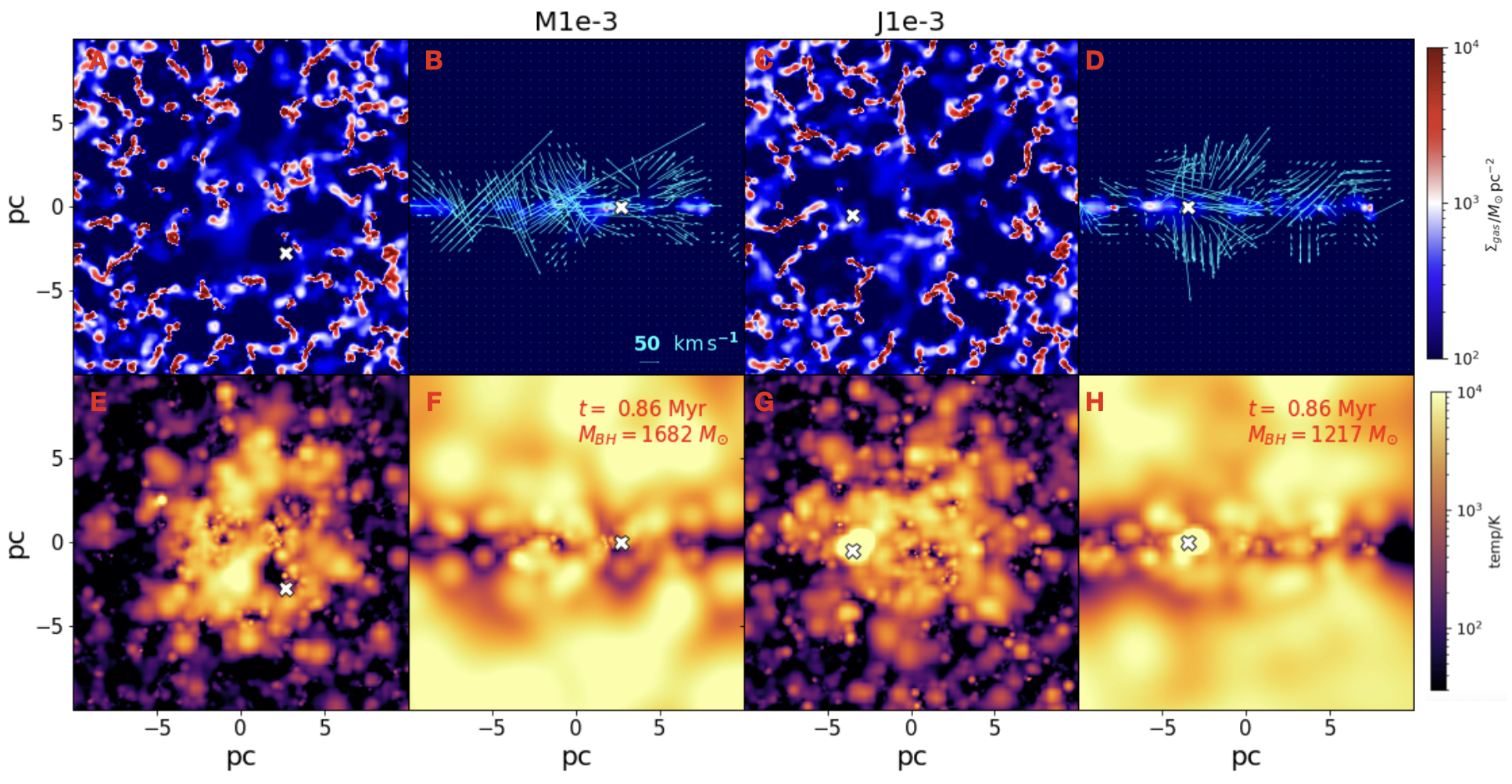}
    \caption{Same as Fig.~\ref{fig:snap1e3}, but at $t=0.86$~Myr.}
    \label{fig:snap2e3}
\end{figure*}

\begin{figure*}   
    \centering
    \includegraphics [scale=0.24]{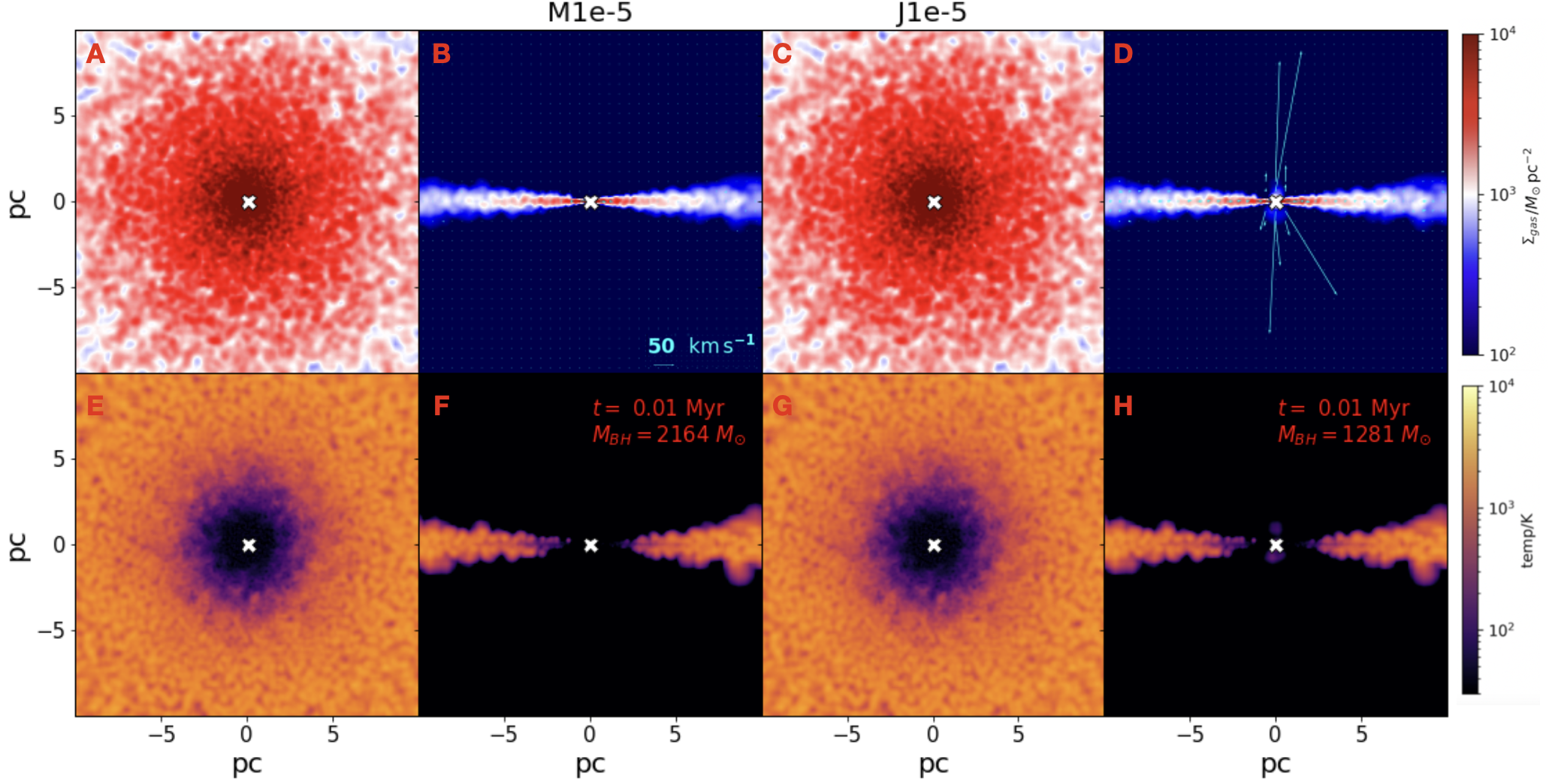}
    \caption{Same as Fig.~\ref{fig:snap1e3}, but for models M1e-5 (panels A, B, E, and F) and J1e-5 (panels C, D, G, and H).}
    \label{fig:snap1e5}
\end{figure*}

\begin{figure*}   
    \centering
    \includegraphics [scale=0.24]{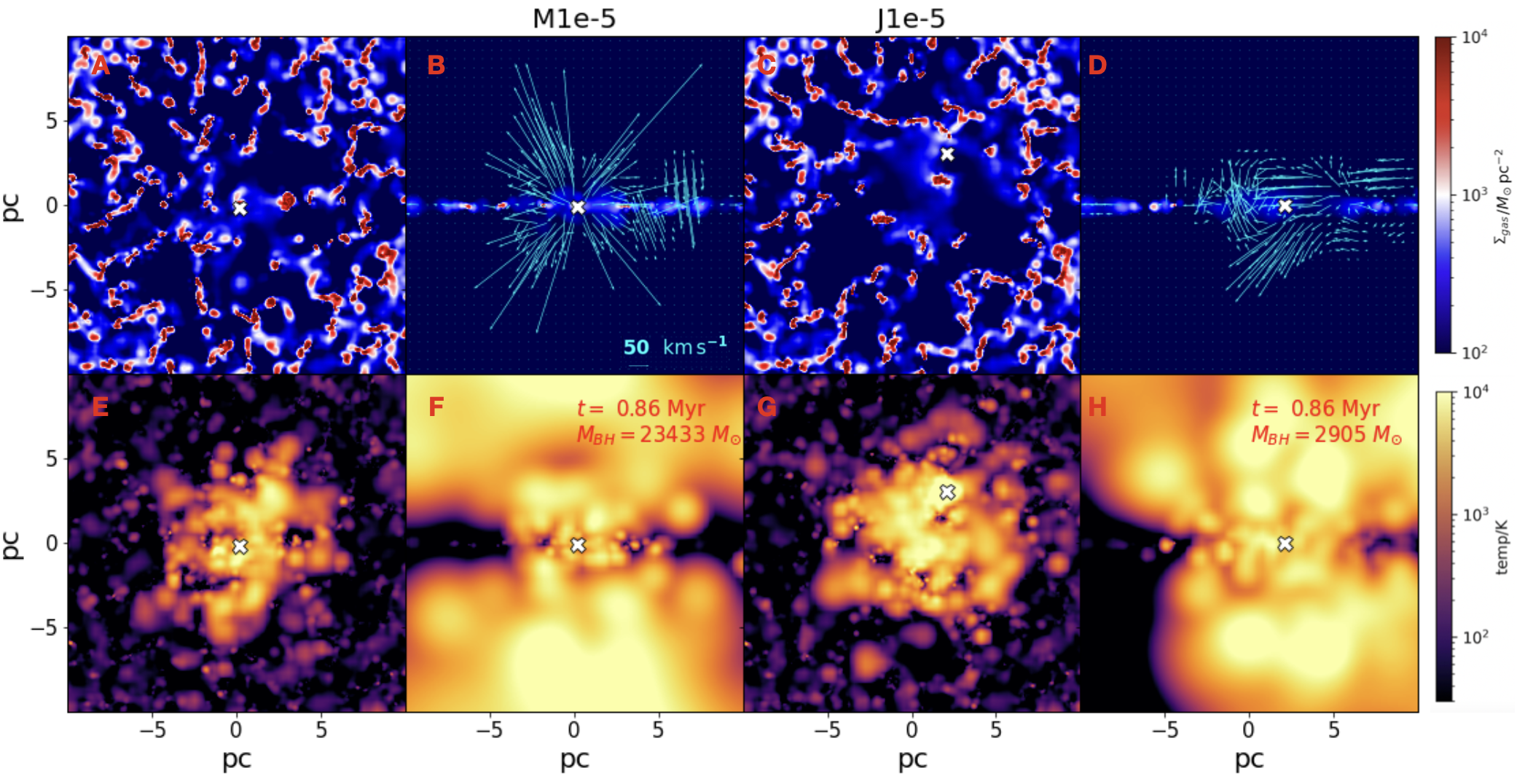}
    \caption{Same as Fig.~\ref{fig:snap1e5}, but at $t=0.86$ Myr.}
    \label{fig:snap2e5}
\end{figure*}

\subsection{Maximal feedback model}\label{subsection:Maximal}

We start by comparing the results of the two modes for a fixed value of $\epsilon_{\rm c} = 10^{-3}$, which is the one commonly adopted in this kind of simulations \citep[e.g.][]{Capelo2015,SouzaLima2017}. In Fig.~\ref{fig:1e3}, we show the radiative efficiency and the Eddington ratio as a function of time. All quantities are averaged over a time interval ${\rm d}t = 0.01$~Myr. In each panel, we show the results of both the Jiang (green-cyan lines in the right-hand panels) and slim disc models (red-yellow lines in the left-hand panels), for a very high value of the spin, $a=0.99$.

When $\epsilon_{\rm c}=10^{-3}$, the gas accretion rate onto the BH is highly variable with time. In both models, BH accretion is highly SE at the beginning of the simulation, with $f_{\rm E} \sim 300$ and 1000 in the Jiang and slim disc models, respectively, and remains so until, at $t\sim 0.2$~Myr, the accretion rate begins to drop below Eddington values (see the lower panels). Between 0.2~Myr and 0.4~Myr, the variability of the gas accretion rate increases and $f_{\rm E}$ shows large oscillations, in some cases varying between $f_{\rm E} \sim 10^3$ and $f_{\rm E} \sim 10^{-2}$ in only a few time-steps. When $t > 0.4$~Myr, the accretion rate is mostly sub-Eddington in both models, with a few isolated episodes of SE accretion. In Section~\ref{subsection:prof}, we show that the decline of the gas accretion rate with time is due to the combined effect of SF, BH feedback, and limited gas mass reservoir. As expected, most of the BH mass growth occurs in the first 0.4~Myr (see also Fig.~\ref{fig:bhevo}), when accretion is mostly SE, and saturates at $t > 0.4$~Myr, with only a few short episodes of significant mass growth before the end of the simulation at $t \sim 1$ Myr. However, there are differences between the two models. In M1e-3, BH mass growth is systematically larger than in J1e-3, and the final BH mass is 1680~M$_{\sun}$ in the slim disc case and 1220~M$_{\sun}$ in the Jiang case. Interestingly, these masses are, respectively, $\sim$ 70 and $\sim$ 20 per cent larger than the mass that the BH would have if it were to accrete at an Eddington-limited rate, with the BH basically not growing in the first Myr. 

These differences can be attributed to the different radiative efficiencies predicted. As shown in the top-right panel, the radiative efficiency in the Jiang model jumps from its lower value, $\epsilon_{\rm r} = 0.01$, at the beginning of the simulation (when $f_{\rm E,J} > 100$, this reference value is shown as a horizontal dashed line in the bottom panel), to its upper value, $\epsilon_{\rm r} = 0.06$, in only a few time-steps of the simulation, and remains more or less constant thereafter. Conversely, in the slim disc model, the radiative efficiency is always smaller than 0.06 in the first 0.2~Myr, with a smooth increase from $10^{-3}$ to 0.06 in the first 0.2~Myr and large oscillations at $0.2 \;{\rm Myr} < t < 0.4$~Myr, when it also exceeds 0.1 in some isolated time-steps. Finally, when $t > 0.4$~Myr, the radiative efficiency in the slim disc model is larger than in the Jiang model (with the exception of a few isolated time-steps) and reaches the asymptotic value of $\sim$ 0.3 (the maximum radiative efficiency for a very high value of $a=0.99$).

The differences of the radiative feedback strengths is also reflected in the time evolution of the temperature and density structure of the inner region of the proto-galaxy, presented in Figs~\ref{fig:snap1e3} and \ref{fig:snap2e3}, at two time-steps of the simulation, $t = 0.01$ and $0.86$~Myr. The odd and even rows show the edge-on and face-on views and the cyan arrows indicate the direction and magnitude of the velocity field. The comparison between the two models shows that at $t = 0.01$ Myr (Fig.~\ref{fig:snap1e3}) the BH has released a larger amount of energy in the Jiang model (panels C, D, G, and H), as a consequence of the larger radiative efficiency, with respect to the slim disc model (panels A, B, E, and F). This leads to a stronger gas outflow and to higher temperatures in the region surrounding the BH (whose position is indicated by a white cross). On the contrary, the differences between the two models, both in the density and temperature structure of the inner region of the proto-galaxy become far less important when $t \geq 0.4$~Myr, as the disc is highly fragmented due to cooling and SF. The {\rm disc} is completely destroyed by $t \sim 1$~Myr, when the BHs are surrounded by low-density hot gas and BH mass growth has already terminated. Fragmentation ought to play an important role in the dynamical evolution  as well as for the growth of the BH. Simulations carried out at various scales, from galactic to nuclear and sub-nuclear, have addressed the effect of a clumpy distribution on the growth and orbital evolution of the lighter secondary companion of a massive BH. For example, \citet{Fiacconi_et_al_2013} and \citet{Tamburello_et_al_2017} have found that clumps  with masses comparable to that of the BH can scatter it out of the disc plane, whereas smaller ones can still lead to perturbations of the orbit and hamper orbital decay. As the mass distribution becomes highly clumpy due to fragmentation in our simulations, the central BH is perturbed gravitationally by massive clumps, which results into sloshing of the central BH, which is reminiscent of the stochastic orbital evolution seen in numerical experiments for massive BH binaries. In a clumpy disc, the density is highly inhomogeneous, and the mass becomes concentrated in the massive star-forming clumps. However, by volume, underdense regions are at least as common. As shown in Fig.~\ref{fig:snap2e3}, as the BH wanders around the centre, it finds itself often surrounded by an underdense region, which contributes to stifling accretion and growth.

\subsection{Intermediate feedback model}

In this section, we discuss the results of the runs when we adopt a lower coupling efficiency: $\epsilon_{\rm c}=10^{-5}$. The results are illustrated in Fig.~\ref{fig:1e3} (M1e-5 and J1e-5). The first important result is that the lower value of the coupling efficiency reduces the effects of feedback and allows larger gas accretion rates, which remain substantially SE for most of the evolution. The largest difference (with respect to when $\epsilon_{\rm c} = 10^{-3}$), is seen in M1e-5, where the accretion rate is less variable in the first 0.2~Myr and remains highly SE, with $10^2 < f_{\rm E} < 10^6$. Conversely, in the same time interval, in J1e-5, $f_{\rm E}$ rapidly drops from $2 \times 10^3$ to $10^2$. When $t > 0.2$~Myr, in both models $f_{\rm E}$ oscillates between $\sim 10^3$--$10^4$ (respectively in J1e-5 and M1e-5) and $\sim 1$, which correspond to accretion rates varying between 10 and $10^{-3}$~M$_{\sun}$~yr$^{-1}$, with a declining trend in the last 0.3~Myr of the evolution. The resulting BH mass growth is very different in the two models: in the first 0.2~Myr, the BH mass in M1e-5 increases by a factor of $\sim$15 and at the end of the evolution reaches $2.36 \times 10^4$~M$_{\sun}$. Conversely, in J1e-5, the growth is less effective and the final BH mass is 2910~M$_{\sun}$. Although limited, this is still a factor $\sim$3 larger than the BH mass predicted in standard Eddington-limited models. 

The differences between the Jiang and slim disc models are largely due to different strengths of BH feedback. Since BH accretion is always highly SE, the radiative efficiency in M1e-5 is smaller than in J1e-5 ($0.01 < \epsilon_{\rm J} < 0.05$) for most of the evolution (top panels). Similar conclusions obviously apply to the BH feedback parameter, i.e.  $\epsilon_{\rm c} \epsilon_{\rm r}$. 

These differences are also very evident when looking at the density, temperature, and velocity structure of the gas in the inner region of the proto-galaxy (Figs~\ref{fig:snap1e5} and \ref{fig:snap2e5}). Fig.~\ref{fig:snap1e5} shows that, at $t = 0.01$ Myr, the effects of BH-driven outflows are already affecting the gas velocity structure at the centre of the inner region of the proto-galaxy, in the area surrounding the BH in J1e-5. Conversely, in M1e-5, there is no BH-driven outflow. The difference between the two models is less apparent at $t \sim 1$ Myr (Fig.~\ref{fig:snap2e5}), although in J1e-5 the gas has been completely drained out of the inner region of the proto-galaxy and the BH appears to be residing in a lower-density environment compared to M1e-5.

From the comparison between the $\epsilon_{\rm c}=10^{-3}$ and $\epsilon_{\rm c}=10^{-5}$ runs, we conclude that -- depending on the SE model adopted -- there is a strong dependence on $\epsilon_{\rm c}$, with the slim disc models providing a more efficient growth for lower $\epsilon_{\rm c}$. In contrast, the Jiang models seem to be less affected by $\epsilon_{\rm c}$. Moreover, independently of the BH feedback efficiency, after 0.4--0.9~Myr, $f_{\rm E}$ drops below 1, and accretion stops in all models.

\subsection{Minimal feedback model}\label{subsection:Minimal}

In this section, we present the time evolution predicted by the two models with the minimal coupling efficiency of $\epsilon_{\rm c} = 10^{-6}$ (M1e-6 and J1e-6). As expected, the evolution of the two models is very similar and so are the final BH masses, which in both cases reach a value of $\sim$2$ \times 10^4$~M$_{\sun}$ (see Fig.~\ref{fig:bhevo}), similar to the case with zero feedback (see Table~\ref{tab:table1Gas}). With the exception of the initial 0.2~Myr, when the smaller radiative efficiency leads to a larger accretion rate in M1e-6, it appears that BH growth is no longer regulated by BH feedback but rather by SF and by the location (and relative velocity to the gas) of the BH itself.

It is interesting to note that in the slim disc models the largest final BH mass is not that predicted by the minimal BH feedback model (M1e-6) but rather by the moderate BH feedback model (M1e-5). This is probably due to the effects of BH feedback on the SFR, which result in a net positive effect on BH growth. A similar conclusion applies when discussing the results obtained by the slim disc models for the non-spinning BHs, Ma01e-3 and Ma01e-6 (see the Appendix).

\begin{figure}
    \centering
    \includegraphics [scale=0.17]{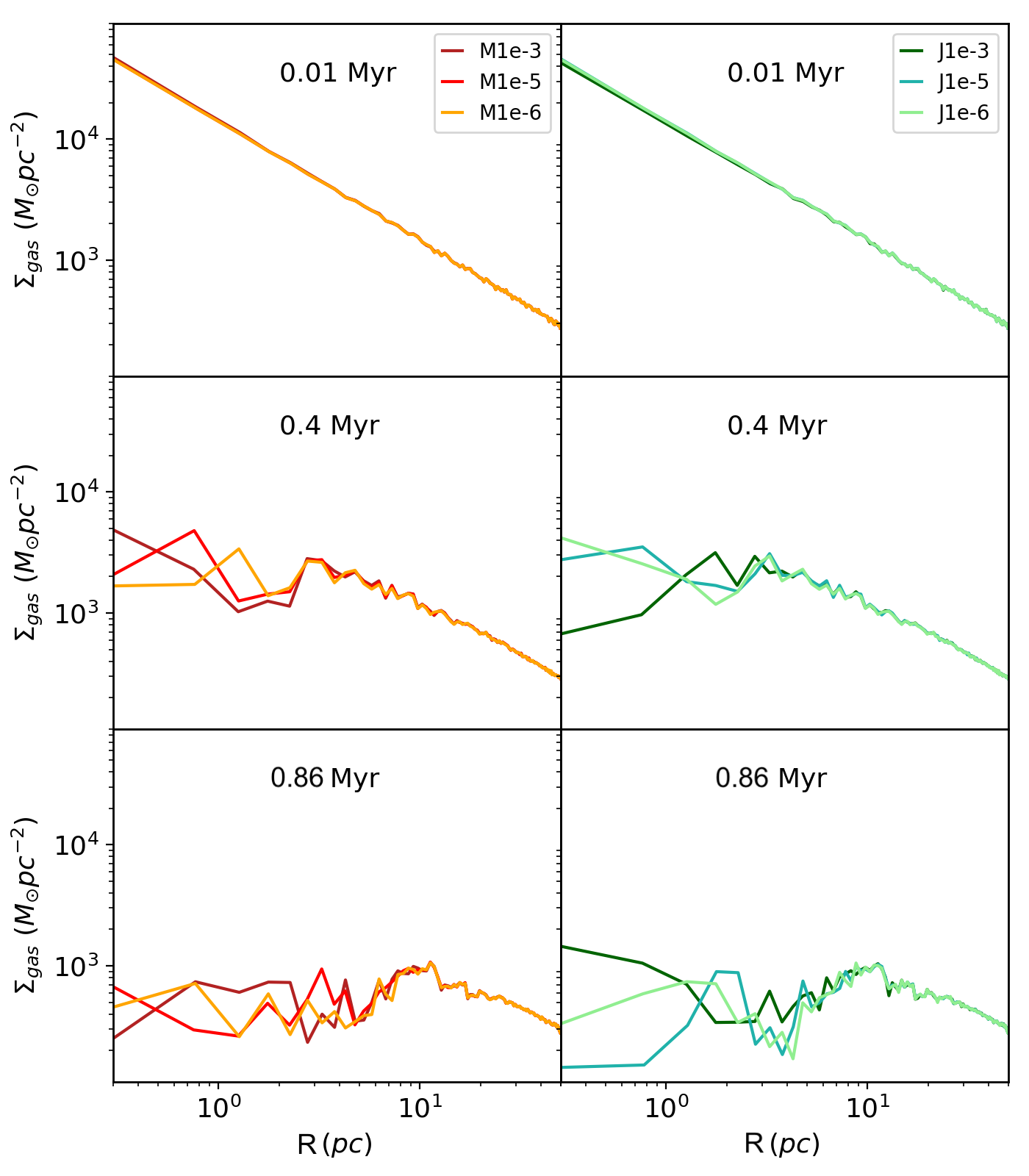}
    \caption{Radial profiles of the gas surface density. The left-hand and right-hand columns show, respectively, the slim disc and Jiang models at different times ($t= 0.01, 0.4$, and 0.86~Myr, from top to bottom) and for three values of the coupling efficiency ($\epsilon_{\rm c}=10^{-3}$, $10^{-5}$, and $10^{-6}$).}
    \label{fig:Sigmadens}
\end{figure}

\begin{figure}
    \centering
    \includegraphics [scale=0.17]{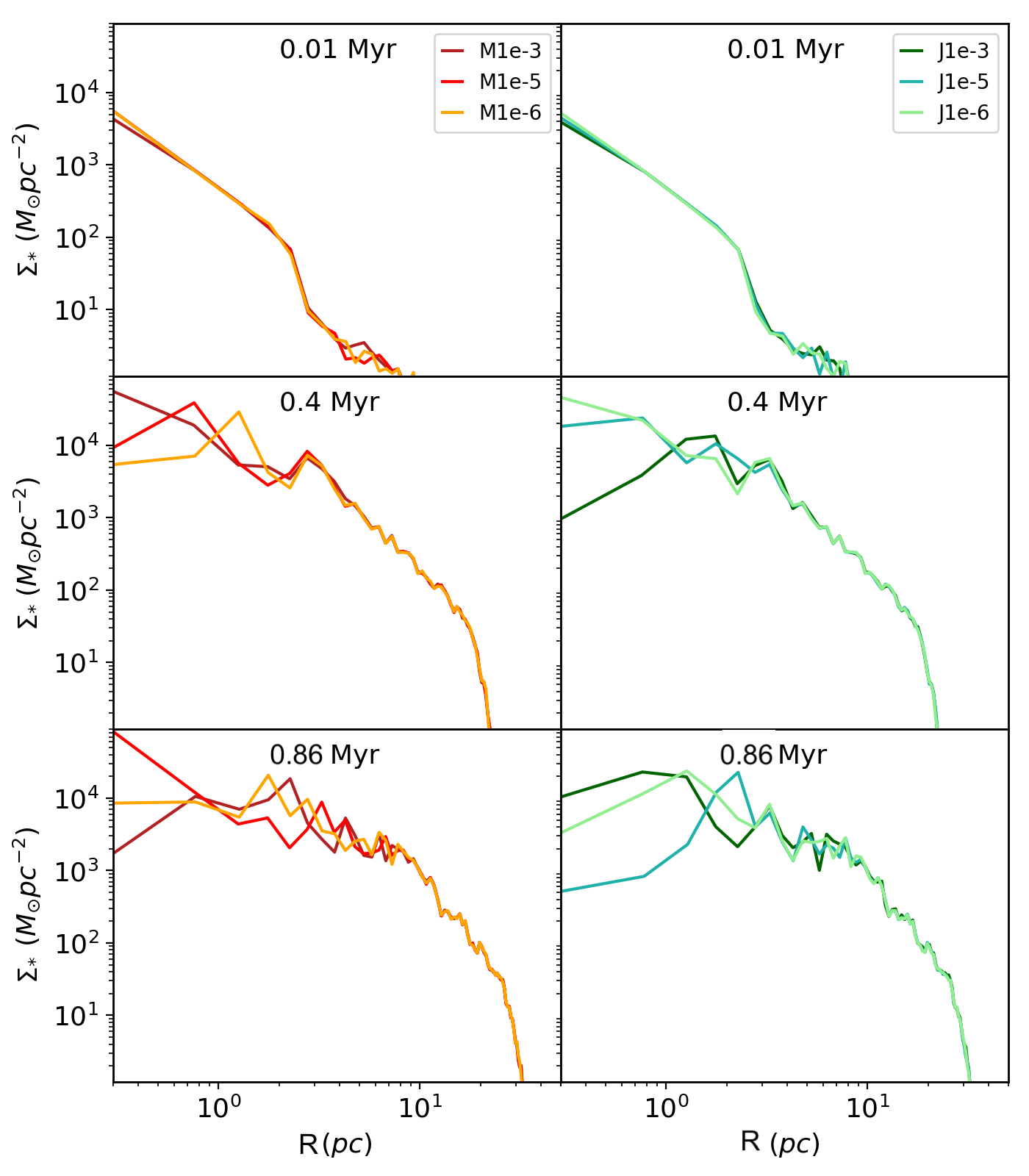}
    \caption{Same as Fig.~\ref{fig:Sigmadens}, but for the stellar surface density.}
    \label{fig:Stardens}
\end{figure}

\begin{figure*}
    \includegraphics [scale=0.5]{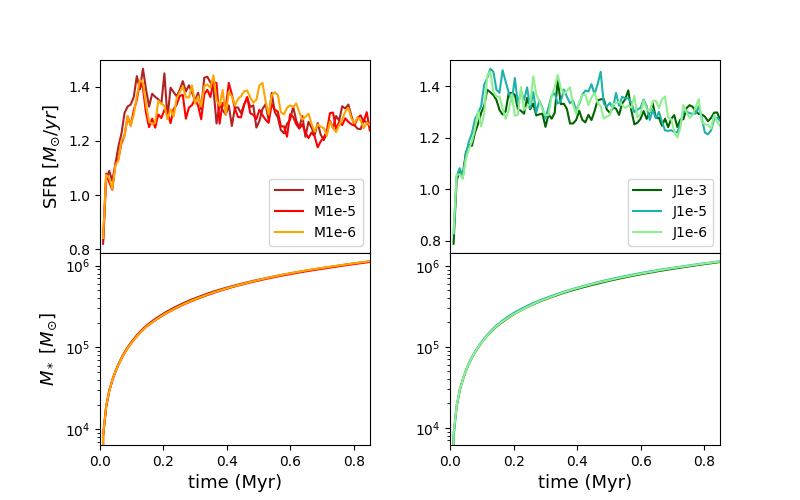}
    \caption{Total SFR (top panels) and total stellar mass (bottom panels) as a function of time in the slim disc (left-hand panels) and Jiang (right-hand panels) models.}
    \label{fig:MstarSFR}
\end{figure*}

\subsection{Disc evolution}\label{subsection:prof}

In Figs~\ref{fig:Sigmadens} and \ref{fig:Stardens}, we show, respectively, the radial profiles of the gas and stellar surface density, for models M1e-3, M1e-5, and M1e-6 (left-hand panels), and J1e-3, J1e-5, and J1e-6 models (right-hand panels) at $t = 0.01$, $0.4$, and $0.9$~Myr. At the beginning of the simulation (0.01~Myr), all models have similar gas and stellar density radial profiles, as expected. As time evolves, we observe a decrease in the gas content within the inner regions, for $r \lesssim [3$--4]~pc at $t = 0.4$~Myr and $r \lesssim [10$--20]~pc at $t = 0.9$~Myr. Outside this central region, the radial profile of the gaseous disc does not change for the entire duration of the simulations. The central gas surface density decreases as a function of time, while the stellar surface density increases. SF starts in the central region and gradually evolves towards the outer regions, with critical consequences on the gas supply that can no longer feed the BH.

In Fig.~\ref{fig:MstarSFR}, we show the total SFR and the total stellar mass as a function of time in all SE models. We observe a rapid increase of the SFR during the first $0.1$~Myr of evolution, and then an oscillating behaviour which stabilizes around 1.3~M$_{\sun}$~yr$^{-1}$. The total stellar mass formed in the disc has the same time evolution in all models, reaching a final value of $M_* \sim 10^6$~M$_{\sun}$, $\sim$20 per cent of the initial gas mass.

In conclusion, our simulations find that different BH feedback efficiencies contribute to evacuating the gas in the region surrounding the BH, within a radius of $\sim$10~pc, and then influence the BH growth during the first $ < 1$~Myr. However, globally the gas budget is consumed by SF. This, together with the BH wandering in low-density regions (and having a relatively high velocity with respect to the gas) strongly suppresses and ultimately stops gas accretion on to the BH when $t > 1$~Myr. Hence, our simulations show that SE accretion can not be sustained on time-scales longer than $\sim$1~Myr \citep[consistent with][]{lupi2016growing,regan2019super}, unless an external gas supply is provided to the system, an event that we cannot describe in our simulations, where the systems are assumed to evolve in isolation, with no gas inflows or mergers.

\begin{figure*}
    \includegraphics [scale=0.6]{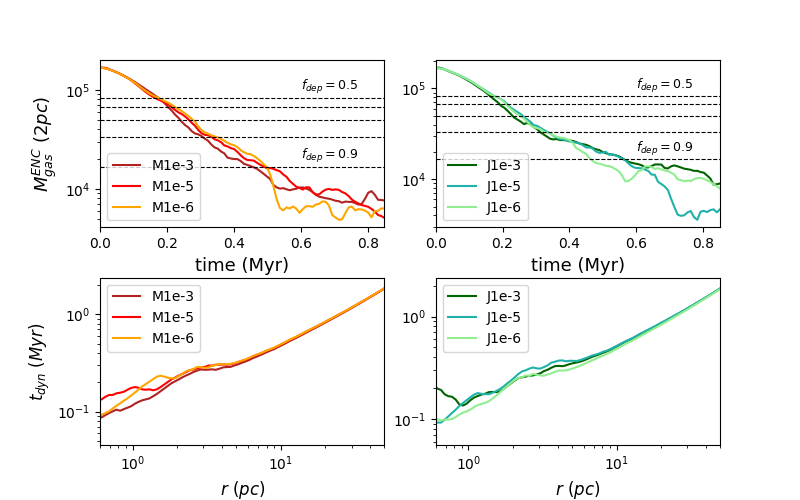}
    \caption{Top panels show the mass enclosed within a sphere of radius $r_*=2$ pc centred on the BH, as a function of time, for the slim disc (left-hand panels) and Jiang (right-hand panels) models. The horizontal black dashed lines indicate different depletion factors: $f_{\rm dep} = 0.5, 0.6, 0.7, 0.8$, and 0.9 (from top to bottom). The bottom panels show the dynamical time-scale as a function of the radial distance from the centre for the slim disc (left-hand panels) and Jiang (right-hand panels) models evaluated at their corresponding $t_{\rm dep}$ (see text).}
    \label{fig:Menc}
\end{figure*}

\subsection{Gas depletion}\label{subsection:depl}

Both BH accretion and SF are strongly dependent on the gas content of the disc. The BH stops growing in less than 1~Myr, because the gas in the central region is depleted by SF (and the BH, due to interactions with clumps, wanders in low-density regions). However, SF itself stops once the gas has been entirely consumed, after $\sim$4~Myr from the beginning of the simulations. Hence, SF and BH accretion are affected by gas depletion on different time- and spatial scales.

In the top panels of Fig.~\ref{fig:Menc}, we show the time evolution of the gas mass enclosed within a sphere of radius $r_*=2$~pc centred on the BH for the slim disc (left-hand column) and Jiang (right-hand column) models. The horizontal dashed lines indicate different values of the depletion factor, defined as

\begin{equation}
f_{\rm dep}(r_*,t) = 1 - M_{\rm gas}(r<r_*, t)/M_{\rm gas}(r<r_*, t_0),
\end{equation}

\noindent where $t_0 = 0$.

It is evident that, independently of the value of the coupling efficiency, the mass enclosed within the inner 2~pc is reduced by a factor 0.5 in only $\sim$0.2~Myr, both in the Jiang and slim disc models. The differences between the models are more evident for $f_{\rm dep} = 0.9$: this condition occurs at $\sim$0.5~Myr in models with lower radiative efficiency in the SE regime, such as the slim disc model, and at $\sim$0.6~Myr in more efficient models, such as the Jiang model. When $\epsilon_{\rm c} = 10^{-6}$, both models are 90-per-cent gas depleted at $\sim$0.5~Myr, because BH feedback is negligible in this case.

Since BH growth depends on the gas content within the central region, we define the depletion time, $t_{\rm dep}$, as the time at which the gas mass within $r_* = 2$ pc is 90-per-cent depleted and we compute the dynamical time of the disc as a function of radius in each simulation at its corresponding $t_{\rm dep}$. The results are shown in the bottom panels of Fig.~\ref{fig:Menc}. When $t = t_{\rm dep}$, for both models the dynamical time at a distance of 50~pc from the centre is $\sim$2~Myr. This represents a rough estimate of the time needed for gravitational torques and dynamical instabilities to drive the gas from the outer regions of the disc towards the inner 2 pc region, down to the gravitational radius of the BH. Hence, we can take this time-scale as the minimum time interval between two subsequent SE accretion episodes.

\section{Discussion}\label{section:SEdiscussion}

Our results are in agreement with previous works that follow SE growth of BH seeds in hydrodynamical simulations. \citet{regan2019super} found that, assuming the same radiative feedback model as in \citet{madau2014} and considering, in addition, the effect of mechanical feedback in the forms of bipolar jets, a $10^4$~M$_{\sun}$ BH can at most enhance its mass by a few per cent in less than 1~Myr. In their case, this is primarily due to the local effect of the bipolar jets, which evacuate the gas from the immediate vicinity of the BH (on scales of $\sim$0.1~pc), suppressing BH growth and reducing the effective accretion rate to a factor of a few below the Eddington rate. Interestingly, however, they find that -- after approximately one dynamical time -- the gas falls back to the centre, powering again a short phase of SE accretion.

These findings were confirmed by the recent work of \citet{massonneau2022supereddington}, who investigated, by means of a high-resolution hydrodynamical simulation, SE accretion on to a $10^6$~M$_{\sun}$ BH in an isolated galaxy hosted by a DM halo of $10^{11}$~M$_{\sun}$ at $z = 4$. They adopted the slim disc model to describe the radiative efficiency (assuming a BH spin of $a = 0.7$) and included both radiative and mechanical feedback from the BH. They found that, during SE accretion, powerful jets can eject gas from the centre up to kpc scales. Although this mechanical feedback does not impact SF, it effectively regulates BH growth, which is strongly intermittent. The final BH mass is therefore strongly dependent on the adopted SE jet feedback efficiency, with lower values allowing for more frequent SE-accretion episodes and a faster BH mass growth. When neglecting the effects of mechanical feedback, \citet{massonneau2022supereddington} found that SE accretion can slightly increase the mass of the BH relative to the standard Eddington-limited case, reaching a final value of $\sim$10$^7$~M$_{\sun}$ in 80~Myr. However, a more refined treatment of radiation and mechanical feedback is needed, and we need to describe the global properties of outflows in a self-consistent way. In this sense, high-resolution 2D RHD simulations from \citet[][]{hu2022long}, provide useful and simple formulae to follow the BH accretion rate in SE regimes, well suitable for galactic-scale simulations.

Our work follows an approach similar to the study of \citet{lupi2016growing}, who investigated the growth of stellar-mass BHs in the CND of an isolated gas-rich, high-redshift galaxy, using the radiatively inefficient feedback model of \citet{madau2014} and neglecting the effects of mechanical feedback by bipolar jets. They found that stellar-mass BHs can occasionally bind to very high density gas clumps that arise from the fragmentation of the surrounding gas. As a result, they can increase their mass by up to three orders of magnitude in less than 1~Myr, beyond which BH growth is suppressed by gas consumption due to SF. It is important to stress that, in their idealised simulations, \citet{lupi2016growing} start from an initial population of equal-mass BHs (with two possible values: 20 or 100~M$_{\sun}$) that are the remnants of previous SF occurred in the CND and that are initially randomly distributed within the disc. The binding between dense gas clumps and the BHs generally occurs when the BHs are found in the disc. When the BHs become massive and are dragged to the centre by dynamical friction, these encounters do not occur. We stress that our medium-weight BHs are placed at the centre of a halo where SF has not occurred yet. In this central region of the disc, BH feedback triggers stronger outflows with respect to those in \citet{lupi2016growing} (who also cap the BH accretion to $500 \,\dot{m}_{\rm E,M}$, whereas we do not impose any limit), and in this sense can contribute to evacuate the innermost region, preventing further accretion. This may partly explain the different findings with respect to \citet{lupi2016growing}, although we note that their coupling efficiency is at least two orders of magnitude higher than ours. Since in \citet{sassano2021light} we find that SF in the same host may have preceded the formation of a medium-weight BH seed, in a future investigation we may consider adding a population of lighter, stellar-mass BHs, similarly to \citet{lupi2016growing}, following their competitive growth and eventually their interactions with the central medium-weight seed.

\subsection{A simple model extrapolation}\label{subsection:simplemodel}

In all of the above studies, and similarly to the present one,  BH growth can only be followed for a limited period of time, due to the idealised conditions adopted in the simulations. Very likely, on longer time-scales, BH growth will be determined by the subsequent evolution of the host galaxy, whose gas content in the central regions could be replenished by gas accretion from large-scale filaments or by merging with other galaxies \citep{pezzulli2016, pezzulli2017sustainable}. However, due to the extensive range of spatial scales required to investigate the efficiency of gas accretion (from galaxy scales down to the BH accretion disc), there are still considerable uncertainties on the role of SE accretion for the formation of $z = 6$ SMBHs.

\begin{table}
\centering
\caption{Average BH mass growth rate in a time interval $\Delta t = 1$~Myr predicted by different runs.}
\begin{tabular}{|c|c|}
\hline 
  Model &  $\Delta M_{\rm BH}/\Delta t$ $[10^3$~M$_{\sun}$Myr$^{-1}$] \T \B \\
  \hline
  M1e-3 &  $0.68 $ \T \B \\
  M1e-5 &   $22.6 $ \T \B \\
  M1e-6 &   $ 19.3 $ \T \B \\
  \hline 
  J1e-3 &  $0.22$ \T \B \\
  J1e-5 &  $1.91$ \T \B \\
  J1e-6 &  $ 17.6$ \T \B \\
\hline
  STD1e-3 &  $0.004 $ \T \B \\
  STD1e-5 &  $0.009 $ \T \B \\
\hline
\end{tabular}
\label{tab:tableDM}
\end{table}

\begin{figure*}
    \includegraphics [scale=0.42]{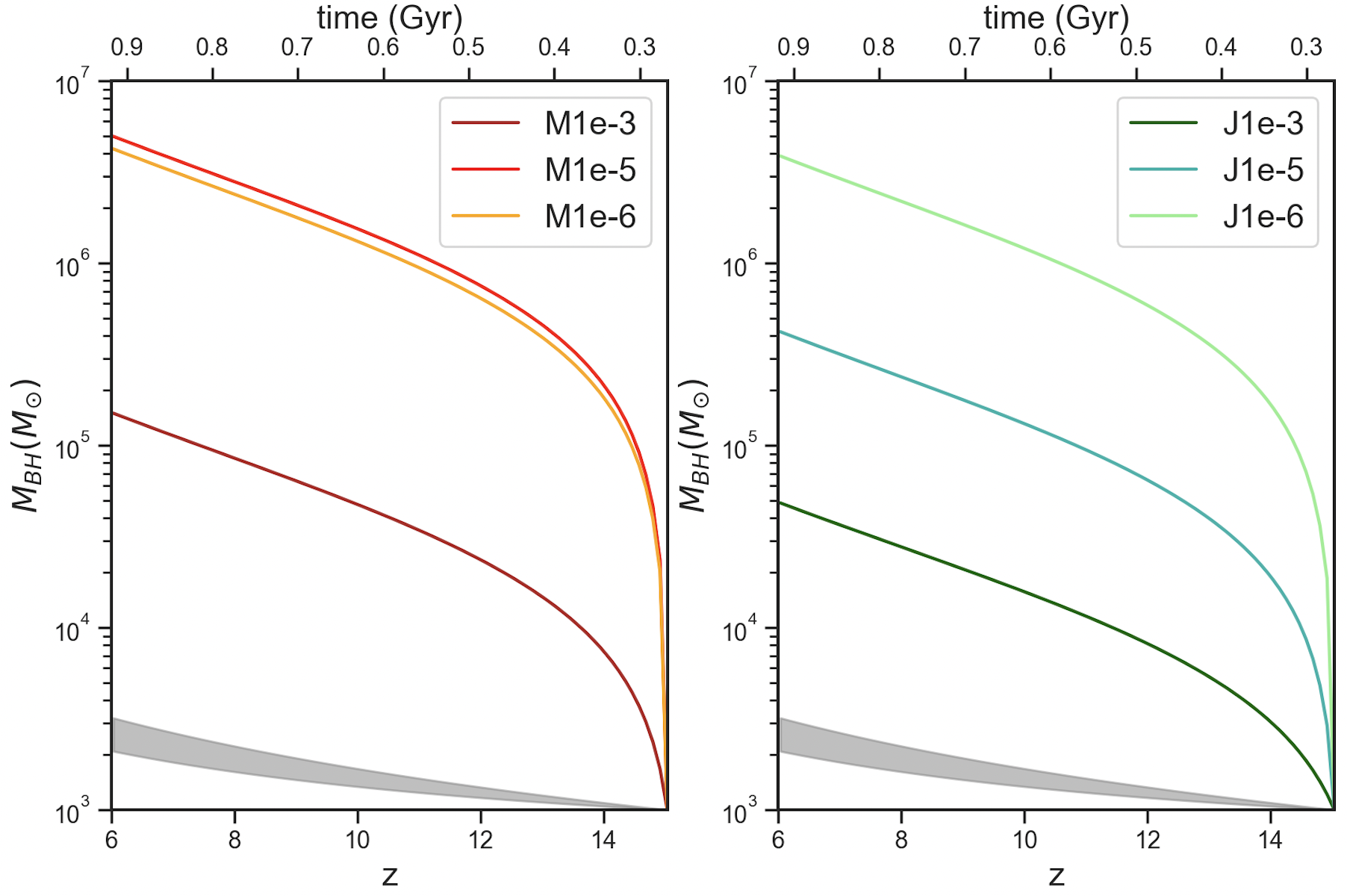}
    \caption{Extrapolated BH mass evolution from $z=15$  to $z = 6$, in the six different SE accretion models, slim disc (right-hand panel) and Jiang (left-hand panel) with different coupling efficiencies, assuming a dynamical time of $t_{\rm dyn}=2$~Myr (see text). The grey-shaded regions indicate the values predicted by the two Eddington-capped simulations.
    }
    \label{fig:estr2}
    \hspace{-0.55 cm}
    \includegraphics [scale=0.35]{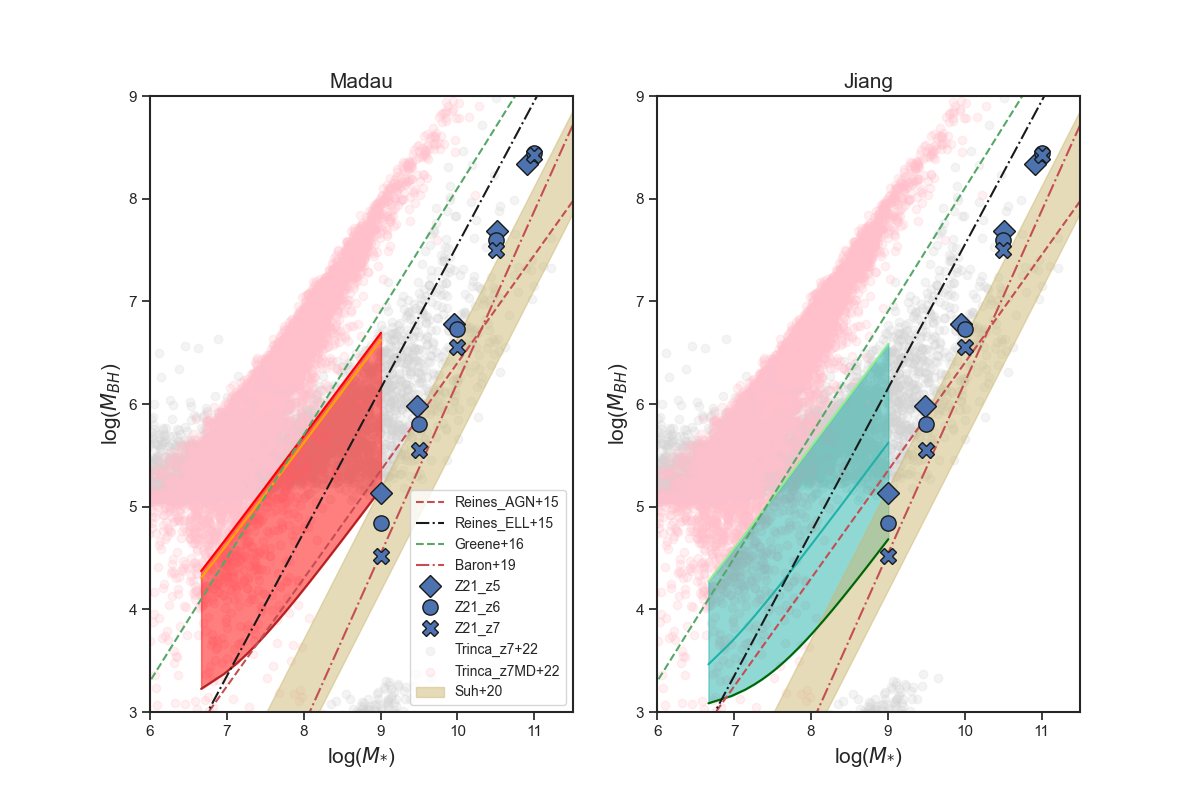}
    \caption{BH mass as a function of the total stellar mass in the cosmological extrapolation of the slim disc (left panel) and Jiang (right panel) models, assuming a dynamical time of $t_{\rm dyn}=2$~Myr (see text). Red and green colored regions between continuous lines indicate the minimum and maximum mass range in the slim disc and Jiang SE accretion models, considering different coupling efficiencies (solid colored lines, with the same color code adopted in Fig.~\ref{fig:estr2}). Dashed lines and points indicate $M_{\rm BH}-M_*$ scaling relation estimated in the local Universe by \citet{greene2016megamaser} (green dashed line), \citet{reines2015relations} (dot-dashed red and black lines), \citet{baron2019black} (red short-dashed line), and \citet{suh2020no} (yellow shaded region). The blue points represent the empirical model of \citet{zhang2021trinity} at  $z=7$, $z=6$, and $z=5$ (crosses, circles, and squares, respectively). We also show the results of the semi-analytical model \textsc{cat} at $6 \leq z \leq 7$: the grey points represent their reference, Eddington-limited model and the pink points their SE merger-driven model \citep{trinca2022low}.}
    
    \label{fig:scaling2}
\end{figure*}

For this reason, we conclude our analysis with an attempt to explore the cosmological evolution of our simulated galaxy, from $z \sim 15$ down to $z \sim 6$. We assume that subsequent episodes of major merger events could rejuvenate the system, promoting gas replenishment in the inner region of the proto-galaxy and BH growth. Each major merger event is supposed to trigger a short ($\sim$1~Myr) phase of SE accretion, following a BH mass growth rate as predicted by our idealised simulations. For each of these, Table~\ref{tab:tableDM} reports the average BH mass growth rate, ${ \Delta M_{\rm BH}/\Delta t}$, over a time-scale of $\Delta t = 1$ Myr. We note that, for all the models, the total stellar mass formed in 4~Myr is $M_* = 4.6 \times 10^6$~M$_{\sun}$.\footnote{As explained in Section \ref{subsection:prof}, even when BH accretion has already halted, SF continues in the external regions of the disc until the gas is completely consumed, on time-scales of $\sim$4~Myr.} We therefore estimate the average SFR over a time-scale of $\Delta t = 4$~Myr as $M_*/\Delta t = 1.15$~M$_{\sun}$~yr$^{-1}$. 

These values are used to extrapolate the final BH and stellar masses at $z \sim 6$. To estimate the maximum number of SE accretion episodes, ${\cal N}_{\rm max}$, we divide the cosmic time between $z=15$ and $z=6$, $\Delta t_{\rm H} = 0.662$~Gyr, by the dynamical time of the galaxy, $t_{\rm dyn}$. We interpret this as the minimum time required to drive new fuel towards the nuclear region, as it would be predicted if gravitational torques and dynamical instabilities were responsible for driving the gas from the outer regions of the disc towards the gravitational radius of the BH. Following the discussion in Section~\ref{subsection:depl}, we find a dynamical time-scale  of $t_{\rm dyn}=2$~Myr, which leads to a number of SE events to ${\cal N}_{\rm max}\sim 220$. The results are illustrated in Fig.~\ref{fig:estr2} that shows the redshift evolution of the BH mass from $z=15$ to $z\sim 6$, for eight of our models: the slim disc models M1e-3, M1e-5, and M1e-6 (left-hand panel), the Jiang models J1e-3, J1e-5, and J1e-6 (right-hand panel), and the Eddington-limited cases STD1e-3 and STD1e-5 (grey-shaded region). We find that the final BH mass depends on the adopted radiative feedback and coupling efficiency, ranging from $5 \times 10^4$~M$_{\sun}$ in J1e-3 up to 4--$5 \times 10^6$~M$_{\sun}$ in M1e-5, M1e-6, and J1e-6. As a comparison, the Eddington-limited models would predict a final BH mass of 2--$3 \times 10^3$~M$_{\sun}$. This simple model extrapolation shows that, even when allowed to increase their mass through short and intermittent SE accretion episodes, medium-weight seeds with mass $\sim$10$^3 $M$_{\sun}$ fail to grow SMBHs similar to those found in quasars at $z \sim 6$. Similar results are found when assuming a dynamical time-scale of $t_{\rm dyn}=10$~Myr between two subsequent SE accretion episodes, as it would be predicted for a proto-galaxy at $z=15$ \citep[e.g.][]{wise2019}. It is important to stress, however, that in all the models the extrapolated stellar mass at $z \sim 6$ is $\sim 10^9$~M$_{\sun}$ (see Fig.~\ref{fig:scaling2}). This value is at least one order of magnitude smaller than the stellar mass of quasar host galaxies estimated from observations (see \citealt{Gallerani2017} for a review on quasar host galaxy properties and \citealt{pensabene2020} for a recent observational study) and predicted by semi-analytical models \citep{valiante2014high, trinca2022low} and high-resolution zoom-in simulations \citep{barai2018, Lupi2019, Valentini2021}. This implies that our simple model extrapolation does not provide a fair representation of the rare overdensities where $z \sim 6$ quasars form and grow, with higher gas accretion rates feeding larger star formation rates and higher gas densities in the central regions. These physical conditions may potentially favour a higher frequency of SE accretion events \citep{inayoshi2016}, up to the point at which the BH becomes massive enough that the mass needed to exceed the Eddington limit is not physically plausible, and the BH continues to grow in an Eddington-limited fashion \citep{pezzulli2016}.

If SE growth can not be efficiently sustained, the formation of $z \sim 6$ SMBHs appears to require more massive seeds, as already suggested by \citet{sassano2021light}, in agreement with previous findings \citep{valiante2016first, valiante2018statistics} and more recently confirmed by means of small-scale 2D RHD simulations coupled to a phenomenological model for the host galaxy evolution \citep{inayoshi2022rapid}.

Yet, their intermittent SE accretion may help medium-weight BH seeds to reach mass values comparable to those observed in the nuclei of present-day low-mass (mostly spiral) galaxies \citep{davis2019}, including our own Milky Way \citep{genzel1996, ghez1998}. As found by \citet[][]{bonoli2016black}, based on cosmological zoom-in simulations, the late-time accretion (at $z < 6$) in galaxies with mass comparable to our grown medium-weight BH seed hosts is rather inefficient. As a consequence, an early phase of SE growth may be required and indeed possible, as our study suggests. 

In Fig.~\ref{fig:scaling2}, we show the relation between the BH mass and the total stellar mass predicted by the SE models in our simple cosmological extrapolation. We find that in this scenario, where SE accretion is short and regulated by SF and BH-clump interactions, subsequent SE-accretion episodes triggered by galaxy mergers lead to an evolution of the BH mass that grows with the total stellar mass along the scaling relation observed at $z = 0$ and predicted at $6 \leq z \leq 7$ by independent models \citep{greene2016megamaser,reines2015relations,baron2019black,suh2020no,zhang2021trinity,trinca2022low}.

It is particularly interesting to compare our predictions with the recent results of \citet{trinca2022low}, who adopted the semi-analytical model \textsc{cat} (Cosmic Archaeology Tool) to explore the evolution of nuclear BHs and their host galaxies at $z > 4$, starting from a population of light ($\sim$100~M$_{\sun}$) and heavy ($\sim$10$^5$~M$_{\sun}$) BH seeds and assuming different BH growth models. In their reference model, shown by the grey points in Figs~\ref{fig:scaling2}, BH growth is assumed to be Eddington-limited, but they also explore an SE scenario (shown by the pink points) wherein BH growth is triggered by galaxy mergers (merger-driven model) and the BH radiative efficiency is described following the slim disc model (with $a = 0.57$). At the low-mass end (which effectively corresponds to $z > 6$) we find that our model predictions are in better agreement with their merger-driven model, as light BH seeds are unable to grow in their reference model and there is a BH mass gap below the mass of heavy BH seeds ($< 10^5$~M$_{\sun}$). However, at the high-mass end, for the same stellar mass, their merger-driven model predicts BH masses that can be up to one order of magnitude larger than our most optimistic model. Hence, our predictions at $z \sim 6$ appear to be in better agreement with their reference, Eddington-limited scenario. This may be an indication that SE accretion is too efficient in their simple model compared to the more sophisticated treatment provided by our hydrodynamical simulations. 

Finally, it is important to stress that we consider an isotropic model for feedback, with important consequences for the outflows. In Fig.~\ref{fig:snap1e3} (panel D), we note that the size of the outflows is relatively broad, and not as narrow as in \citet[][]{jiang2014global}, who found  collimated funnel regions, which help to evacuate the radiation along the polar directions, not obstructing the mass infall along the disc plane. For this reason, we underline that, since we do not have a refined prescription for the radiative transport able to take into account the presence of collimated jets which could reduce the size of the outflows, we are still underestimating the SE mass accretion phase and then the BH mass. 

\section{Conclusions}\label{section:conclusions}

In this work, we investigate SE accretion onto a BH seed of $10^{3}$~M$_{\sun}$ by simulating the inner $\sim$ 50~pc of a metal-poor, gas-rich galaxy at $z = 15$, using the SPH $N$-body code \textsc{gasoline2}. 

We explore two different feedback models to test their effects on the early mass growth of BH seeds. In the first model, we adopted the prescriptions of \citet[][]{madau2014}, who provided an analytic fit to the numerical solutions of the relativistic slim accretion disc equations of \citet[][]{sadowski2009}. In the second model, we considered the recent results of 3D MHD simulations by \citet[][]{jiang2014global}, who find a more radiatively efficient disc compared to the slim-disc solution. Our model does not include a description of mechanical feedback via bipolar jets and  quantifies the effects of the energy released by the accreting BH in terms of radiative feedback, i.e. by thermally coupling a small fraction of the BH luminosity to the surrounding gas. Here we treat the coupling efficiency as a free parameter and we vary its value in the range [$10^{-3}$--$10^{-6}$].

In our simulations, we find that, for standard values of BH feedback coupling efficiency ($\epsilon_{\rm c} = 10^{-3}$), BH accretion is more efficient in the slim disc models, wherein the effect of BH feedback onto the gas is less pronounced due to the lower radiative efficiencies when $f_{\rm E} \gtrsim 1$. The BH mass growth is on average higher in the slim disc models, especially at earlier evolutionary epochs ($t < 0.2$~Myr), when $f_{\rm E} \gtrsim 10^2$ and the BH mass rapidly grows from the initial 1000~M$_{\sun}$ up to 1680~M$_{\sun}$. In contrast, at the beginning of the simulation, the effect of BH feedback is more pronounced in the Jiang models (and thus $\dot M_{\rm BH}$ and $\dot m$ are on average lower), so that the BH mass grows slowly, reaching only 1220~M$_{\sun}$.

We find that BH growth in the Jiang models is less sensitive to the strength of BH feedback than in the slim disc models. As a result, for intermediate feedback models ($\epsilon_{\rm c} = 10^{-5}$), the final BH mass is one order of magnitude higher in the slim disc model ($2.36 \times 10^4$~M$_{\sun}$) than in the Jiang model ($2.91 \times 10^3$~M$_{\sun}$), while in the minimal feedback models ($\epsilon_{\rm c} = 10^{-6}$), the final BH mass is comparable in the two models, with $M_{\rm BH} \sim 2.4 \times 10^4$~M$_{\sun}$. These minimal feedback models show how the BHs would grow in the absence of BH feedback, when the gas supply is only affected by SF and BH-clump interactions. Even in the intermediate feedback models, however, SE growth leads to a final BH mass that is between 3 (for the Jiang model) to 24 (for the slim disc model) times larger than that predicted by a standard Eddington-limited growth.

Finally, for the slim disc models, we explored the effects of changing the spin of the BH, from the very high value of $a=0.99$ discussed above to the non-spinning case ($a=0$), which is consistent with the Jiang model when $f_{\rm E} \lesssim 1$ (see Fig.~\ref{fig:jiangmadau}). When $\epsilon_{\rm c} = 10^{-6}$, non-spinning BHs follow the same evolution of the previous models, as expected, reaching a final mass of $10^4$~M$_{\sun}$. Since the dependence of the radiative efficiency on BH spin is negligible during the SE accretion phase, even when $\epsilon_{\rm c} = 10^{-3}$ non-spinning BHs initially rapidly grow their mass, reaching in a few time-steps a value that is larger than for the case $a=0.99$. However, this increases the strength of BH feedback, which suppresses gas accretion and the BH grows very mildly there after, reaching a final mass of only 1300~M$_{\sun}$.

All runs show that SF consumes the disc after $\sim$1~Myr. As a result, BH accretion stops. Instead, SF continues at the periphery of the disc up to $\sim$4~Myr, where it starts to decline due to the lack of gas. 

We extrapolate the final BH and stellar masses from $z=15$ to $z\sim6$, assuming subsequent galaxy mergers will replenish the gas reservoir and trigger a new cycle of SE accretion {and star formation}. We find that at most BH seeds would grow to $\sim 10^6$~M$_{\sun}$, comparable to the mass of massive BHs in spiral galaxies such as our own Milky Way, but falling short of the mass of the high-redshift quasars and of their host galaxies.

Although based on highly idealised simulations that target the inner region of a gas-rich high-redshift proto-galaxy, our results may provide an indication on how to model short and intermittent SE accretion on to a medium-weight BH seed in cosmological models that describe the mass assembly history of the first SMBHs \citep{sassano2021light}, to investigate whether short and intermittent SE accretion episodes may provide an attractive route to the formation of the first SMBHs \citep{pezzulli2016}. Similarly, they could be implemented in models that describe the redshift evolution of the BH mass function through cosmic time \citep{trinca2022low}, to assess whether future surveys may be able to discriminate among different BH growth models in the early Universe. 

\section*{Acknowledgements}
We thank the reviewer, Alessandro Lupi, for his insightful and constructive comments, which helped to largely improve the quality of the paper. F.S., R.S., and R.V. acknowledge support from the Amaldi Research Center funded by the MIUR program ``Dipartimento di Eccellenza'' (CUP:B81I18001170001) and from the INFN TEONGRAV specific initiative, and the networking support by the COST Action CA16104. F.S. thanks the University of Zurich for the kind support and hospitality.

\section*{Data Availability Statement}
The data underlying this article will be shared on reasonable request to the corresponding author. The code \textsc{gasoline2} \citep[][]{wadsley2004gasoline,wadsley2017gasoline2} is publicly available at \url{https://gasoline-code.com}. For the analysis, we used the package \textsc{pynbody} \citep[][]{pynbody}, publicly available at \url{https://github.com/pynbody/pynbody}.



\bibliographystyle{mnras}
\bibliography{bibliography} 



\appendix

\section{Slim disc model with a non-spinning BH}

\begin{figure}
    \centering
     \includegraphics [scale=0.4]{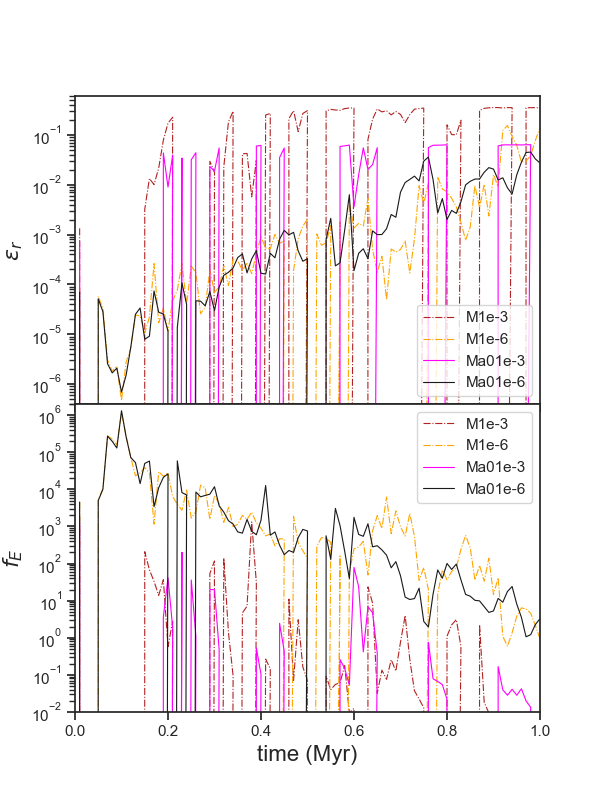}
     \caption{Same as Fig.~\ref{fig:1e3}, but showing runs M1e-6, M1e-3, Ma01e-6, and Ma01e-3.}
    \label{fig:eps2}
\end{figure}

\begin{figure}
    \centering
     \includegraphics [scale=0.19]{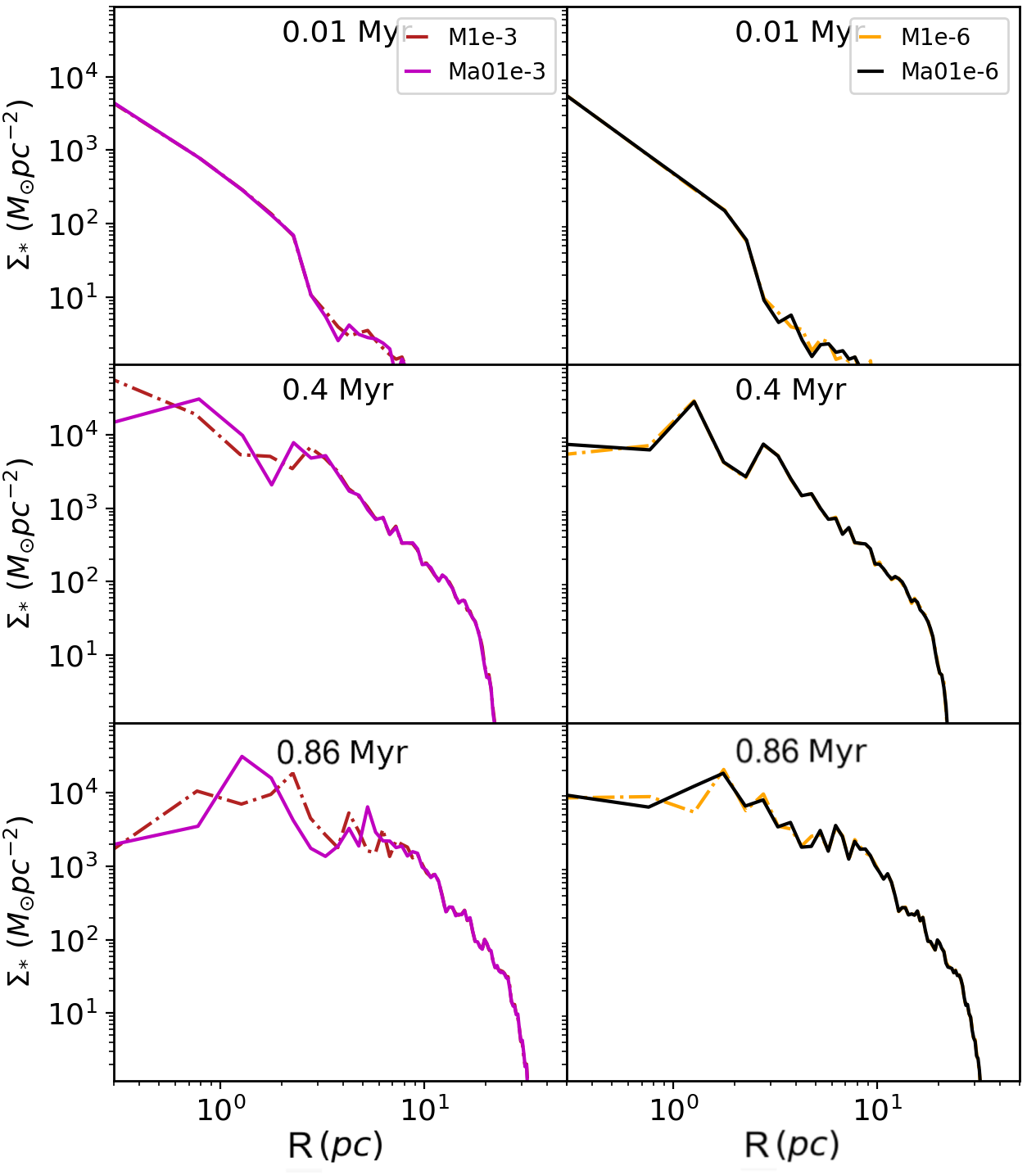}
     \includegraphics [scale=0.19]{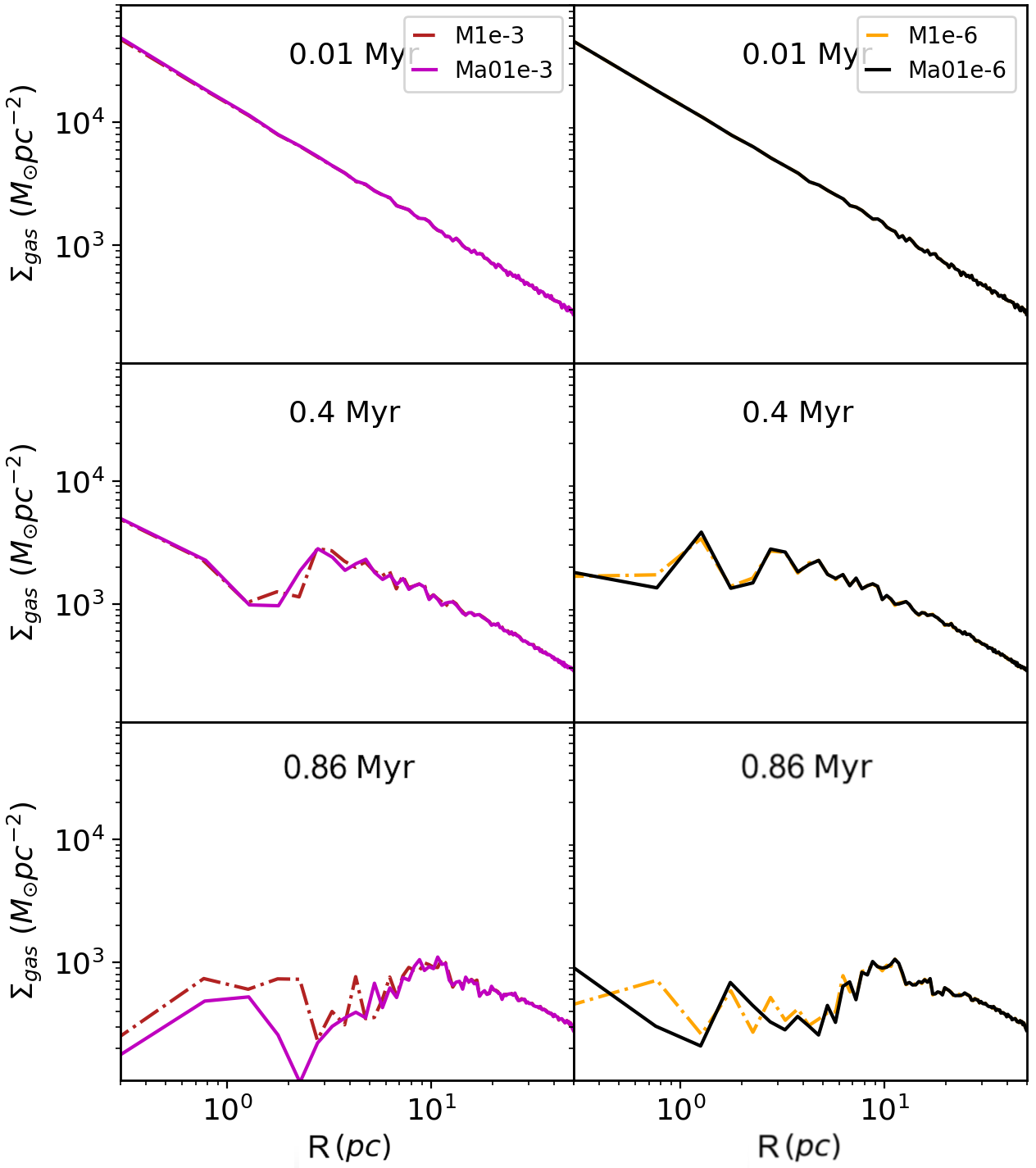}
    \caption{Same as Figs~\ref{fig:Sigmadens} and \ref{fig:Stardens}, but showing the radial profiles of the gas and stellar surface density for runs M1e-6, M1e-3, Ma01e-6, and Ma01e-3.}
    \label{fig:densa0}
\end{figure}

\begin{figure*}
    \centering
    \includegraphics [scale=0.28]{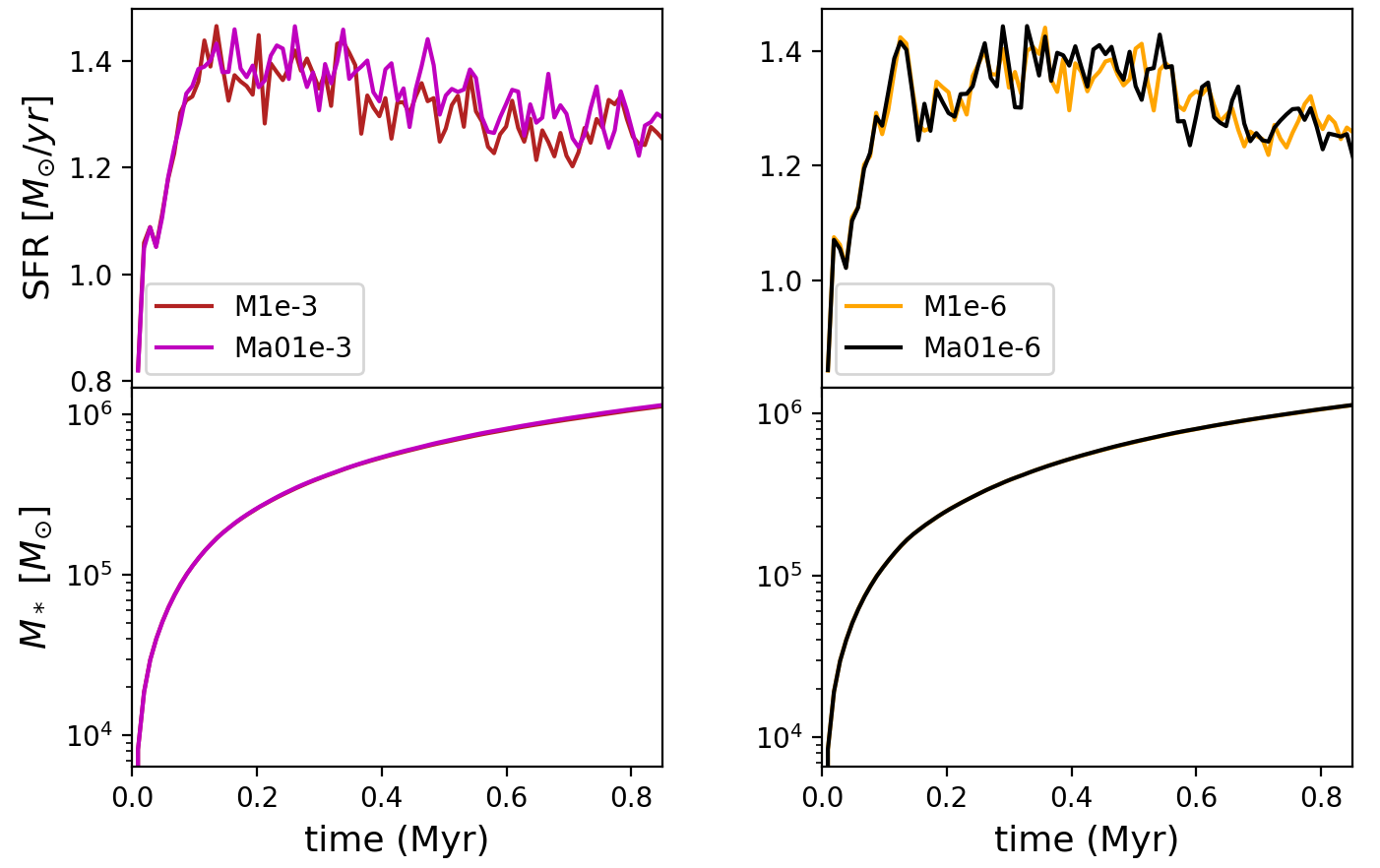}
    \caption{Same as Fig.~\ref{fig:SFMstar} but for runs M1e-6, M1e-3 with $a=0.99$, and Ma01e-6, Ma01e-3 with $a=0$.}
    \label{fig:SFMstar}
\end{figure*}

In this section, we report the results of two additional simulations, Ma01e-3, and Ma01e-6, wherein we assumed $\epsilon_{\rm c} = 10^{-3}$, and $10^{-6}$ and the value of the radiative efficiency predicted by the slim disc model for non-spinning BHs. As shown in Fig.~\ref{fig:jiangmadau}, with this choice the Jiang and slim disc models predict very similar values of $\epsilon_{\rm r}$ when $f_{\rm E} \lesssim 1$.

The magenta and black lines in Fig.~\ref{fig:eps2} show respectively the results of Ma01e-3, and Ma01e-6. Comparing Ma01e-3 with M1e-3 (dot-dashed red line), it is clear that the spin of the BH has a large impact on the time evolution of the gas accretion rate and -- consequently -- on the BH mass growth. At the very beginning of the simulation, the BH mass growth in model Ma01e-3 follows the same behaviour of model M1e-3. Indeed, at the very beginning, gas accretion is highly SE and in this regime $\epsilon_{\rm r}$ has a very small dependence on the spin in the slim disc model. We do not show the gas density and temperature maps of the disc at $t = 0.01$~Myr, which are almost identical to what we find for M1e-3. However, after a few time-steps, the accretion rate in Ma01e-3 drops. We observe substantial feedback from the BH, as in the corresponding Jiang case (J1e-3; green solid lines of Fig.~\ref{fig:1e3}), and also the BH has a comparable final mass (see Fig.~\ref{fig:bhevo}). After this first  episode, the BH evacuates the surrounding region and $\dot m$ drops. Subsequently, the lower feedback at $f_{\rm E} \lesssim 1$ promotes SF, and BH accretion is quenched, similarly to J1e-3. As a result, the time evolution of the BH mass in Ma01e-3 is closer to J1e-3 than to M1e-3, reaching a final value of 1300~M$_{\sun}$. 

It is interesting to compare these findings with the results of Ma01e-6. The upper panel of Fig.~\ref{fig:eps2} shows that, when $\epsilon_{\rm c} = 10^{-6}$, the effects of BH feedback are negligible and -- despite the different values of $\epsilon_{\rm r}$ -- the three SE models with $\epsilon_{\rm c} = 10^{-6}$ (M1e-6, J1e-6, and Ma01e-6) follow a very similar time evolution, leading to similar final BH masses ($2 \times 10^4$~M$_{\sun}$). We observe that BH feedback does not play a role and, after $0.4$~Myr, the disc configuration is the same as in M1e-6. The dependence on the adopted BH spin of the time evolution of gas and stellar density radial profiles predicted in the slim disc models is presented in Fig.~\ref{fig:densa0}, for $\epsilon_{\rm c} = 10^{-3}$ (left-hand panels) and $10^{-6}$ (right-hand panels). As expected, the dependence on BH spin is negligible when $\epsilon_{\rm c} = 10^{-6}$. However, when $\epsilon_{\rm c} = 10^{-3}$, we find that the two models show differences in the stellar density profile (when $t \gtrsim 0.4$~Myr) and in the gas density profile (when $t > 0.4$~Myr) in the inner region of the disc, at $r < 2$--3~pc. These differences are reflected in the time evolution of the global SFR, which is systematically larger in Ma01e-3 than in M1e-3 beyond $t \sim 0.1$~Myr, while it has a similar trend in models M1e-6 and Ma01e-6. In all cases, the differences do not reflect in the time evolution of the total stellar mass, which appears to be insensitive to BH feedback (see Fig.~\ref{fig:SFMstar}).


\bsp	
\label{lastpage}
\end{document}